\documentclass[10pt,letterpaper]{article}
\usepackage[top=0.85in,left=2.75in,footskip=0.75in]{geometry}

\usepackage{amsmath,amssymb}
\usepackage{dsfont} 

\usepackage{changepage}

\usepackage[utf8x]{inputenc}

\usepackage{textcomp,marvosym}

\usepackage{cite}

\usepackage{nameref,hyperref}

\usepackage[right]{lineno}

\usepackage{microtype}
\DisableLigatures[f]{encoding = *, family = * }

\usepackage[table]{xcolor}

\usepackage{array}

\newcolumntype{+}{!{\vrule width 2pt}}

\newlength\savedwidth

\raggedright
\usepackage[aboveskip=1pt,labelfont=bf,labelsep=period,justification=raggedright,singlelinecheck=off]{caption}

\makeatletter
\renewcommand{\@biblabel}[1]{\quad#1.}
\makeatother

\usepackage{lastpage,fancyhdr,graphicx}
\usepackage{epstopdf}
\fancyheadoffset[L]{2.25in}
\fancyfootoffset[L]{2.25in}
\newtheorem{theorem}{Theorem}
\newtheorem{definition}{Definition}
\newtheorem{lemma}{Lemma}
\newtheorem{remark}{Remark}
\newtheorem{property}{Property}

\begin{document}
\vspace*{0.2in}

\begin{flushleft}
{\Large
\textbf\newline{A modified Susceptible-Infected-Recovered model for observed under-reported incidence data} 
}
\newline
\\
Imelda Trejo\textsuperscript{1* \textpilcrow},
Nicolas Hengartner\textsuperscript{1 \textpilcrow}\\
\bigskip
\textbf{1} Theoretical Biology and Biophysics  Group, Los Alamos National Laboratory, Los Alamos, Nuevo Mexico, United States of America
\\
\bigskip
\textpilcrow These authors contributed equally to this work.


* imelda@lanl.gov

\end{flushleft}
\section*{Abstract}
Fitting Susceptible-Infected-Recovered (SIR) models to incidence data is problematic
when not all infected individuals are reported. 
Assuming an underlying SIR model with general but known distribution 
for the time to recovery, this paper derives the implied differential-integral 
equations for observed incidence data
when a fixed fraction of newly infected individuals are not observed.  The 
parameters of the resulting system of differential equations are identifiable.
Using these differential equations, we develop a stochastic model for
the conditional distribution of current disease incidence given the entire past history
of reported cases.
We estimate the model parameters using Bayesian  Markov Chain Monte-Carlo sampling
of the posterior distribution. 
We use our model to estimate the transmission rate and fraction of asymptomatic individuals
for the current Coronavirus 2019 outbreak in eight 
American Countries: the United States of America, Brazil, Mexico, Argentina, Chile, Colombia, Peru, and Panama, from January 2020 to May 2021.
Our analysis reveals that consistently, about
40-60\% of  the infections
were not observed in the American outbreaks.
The two exception are Mexico and Peru, with acute under-reporting in Mexico.

\section*{Author summary}

Quantifying the lethality and infectiousness of emerging diseases such as the coronavirus disease (COVID-19) pandemic is challenging because of under-reported cases of the disease.
Under-reporting can be attributed to the presence of sub-clinical infections, asymptomatic individuals and lack of systematic testing. 
We develop an extension of standard epidemiological models to describe the temporal observed dynamics of  infectious diseases and
to estimate the under-reporting from incidence data. The extended model shows that fitting SIR-type models  directly to incidence data will underestimate the true infectiousness of the disease. Therefore, failing to account for the under-reporting will underestimate the severity of the outbreak, possibly leading decision makers to call the epidemic under control prematurely. Additionally, 
we present a novel stochastic framework to estimate the conditional expectation of incidence  given the past observed cases and to provide a likelihood function for the epidemic model parameters: the transmission rate and the fraction of under-reported cases.  The stochastic model is able to track the various complex COVID-19 incidences with confidence bounds.


\section*{Introduction}

Susceptible-Infected-Recovered (SIR) models, introduced by Kermack and McKendrick and further developed by Wilson and Worcester \cite{Kermack27,Wilson45},
have been extensively used to describe the temporal dynamics 
of infectious disease outbreaks \cite{Anderson92,Brauer12,Diekmann00}.
They have also been widely used 
to estimate the disease transmission rate by fitting the models to observed
incidence data \cite{Chowell04,Chowell06,chowell07}, such as time series of daily or weekly reported number 
of new cases provided by \cite{CDC,JHU,PAHO,WHO}, for example.  
Implicit in all these model fittings is the assumption that all the infected individuals have been observed.  
Yet that assumption is problematic when disease
incidences are under-reported.  Under-reporting 
of incidence is prevalent in health surveillance of emerging diseases \cite{DelValle18,Lai20}, and 
also occurs when a disease presents a large fraction of 
asymptomatic carriers, e.g., Typhoid fever, Hepatitis B, Epstein-Barr virus \cite{Kalajdzievska11} and Zika \cite{Duffy09}.  
Lack of systematic testing and the presence of sub-clinical patients,
which are prevalent in both Severe Acute Respiratory Syndrome Coronavirus 2 (SARS-CoV-2), the causative agent of the coronavirus disease (COVID-19) pandemic \cite{Doll20,Esb20,Li20,chinaCDC}, and Influenza \cite{Furuya16,Reed09}, also leads to under-counting incidence and death.
Directly fitting an SIR model to raw under-reported incidence
will underestimate the transmission rate 
(see Under-estimation of the transmission rate section).
Therefore, failing to account for the under-reporting will under-estimate
the severity of the outbreak, possibly leading decision makers to
call the epidemic under control prematurely.

To account for under-reporting in an SIR-type model, Shutt {\em et. al} \cite{Shutt17} propose to split the infected individuals into two: an observed category and an unobserved category.  This is a special case  of  the Distributed Infection (DI) models introduced in \cite{Hyman99}. However, fitting this model to data is problematic since there are no data from the unobserved category. Furthermore, making inferences about DI model parameters is difficult as there are no adequate stochastic model extensions for the DI models,  which implies that there is no analytic expression for the likelihood.  A partial solution of this problem is to use Approximate Bayesian Computations 
as in \cite{Shutt17} or rely on particle filtering \cite{Romero-Severson2020.04.18.20070771}. Finally, we mention two recent approaches to model asymptomatic individuals in SIR-type models:  First  Lopman {\em et. al.} in \cite{Lopman14} model  Norovirus outbreaks using an SEIR model, with E standing for ``exposed'', where the infected would  progress from symptomatic to asymptomatic to immune. Once immune, 
individuals could cycle between immune and asymptomatic infection.  Second, Kalajdzievska {\em et.  al.} in \cite{Kalajdzievska11} 
propose an SIcIR model, with Ic standing for ``infectious carrier'', where infected individuals are separated into asymptomatic and symptomatic groups by a given probability  as they progress from the susceptible group.

The aim of this paper is to present a novel approach to estimate the under-reported from reported
incidence data and apply this methodology to COVID-19 incidences.  The COVID-19 pandemic is a particular example of an infectious disease that poses many challenges in quantifying the under-reported incidence, and hence  estimating its infectiousness \cite{PPR:PPR141121,Li20,San20}, as under-reporting arises from  the presence of sub-clinical infections \cite{chinaCDC,Bai2020}, asymptomatic individuals \cite{Bar20,Rothe20}, and lack of systematic testing \cite{Doll20,Esb20}.  Accordingly, asymptomatic individuals account for 20-70\%  of all the infections \cite{Bar20}. Additionally, early in the China outbreak, before traveling restrictions,  86\% of all infections were not documented \cite{Li20}.

In the development of our methodology,  
we present two innovations:
First, we introduce an alternative to the DI models that 
directly describes the dynamics of the observed under-reported
incidences. 
Specifically, assuming that a constant fraction 
of the newly infected individuals is observed, we derive a set of  integral-differential equations describing the local temporal dynamics of the
observed incidence.  
Second, we use the local dynamics of the observed incidence to propose 
a model for the conditional expectation of new cases,
given the observed past history.  Making additional distributional assumptions,
we obtain a likelihood for the epidemic model parameters: the transmission rate $\beta$,
and the fraction $p$ of observed incidence.  We refer to Bettencourt and Ribeiro \cite{Ruy2008} for an interesting alternative framework that leads to a likelihood for the basic reproduction number 
${\mathcal R}_0$. We show that as the epidemic progresses, 
both of these parameters become identifiable.

\section*{Materials and methods}

\subsection*{Data source}

The time series of the daily number of confirmed COVID-19 cases and total population,  $N$, of the eight analyzed countries, were obtained from World Health Organization (WHO) reports. Both data sets can be freely downloaded online \cite{WHO, WHO2}. We used all available incidence reports up to the present study, which corresponds to the reports  from January 03, 2020 to May 18, 2021.

\subsection*{Model development}
Our epidemic model is developed in three steps. First, we extend a generalized SIR model to describe the dynamics of the observed (under-counted) infections.  Second, we introduce a local version of that SIR model to describel the evolution of the epidemic in a series of observational time windows given the past time serie of observed incidences.  This more flexible model is used to compute the conditional expectation of current  observed incidence given the past history. Third, we develop a computationally tractable approximation for the conditional  expectation to speed up Monte-Carlo Markov Chain (MCMC) inferences of our model parameters.

\subsubsection*{Generalized SIR model}

Classical mass-action epidemic models, such as the SIR models, are simple yet useful mathematical descriptions of the temporal dynamics of disease outbreak \cite{Anderson92,Brauer12,Diekmann00}. These models describe the temporal evolution of the number of susceptible $S(t)$, infected $I(t)$ and recovered $R(t)$ individuals in a population of fixed size $N=S(t)+I(t)+R(t)$. We model their dynamics through the set of integral-differential equations \cite{Hethcote80}:
\begin{eqnarray}
S^\prime(t)&=& -\frac{\beta}{N} S(t)I(t) \label{eq:Aa}\\
I(t)&=& \int_0^t \left(-S^\prime(u)\right)(1-F(t-u)) du + I(0)(1-F(t)) \label{eq:Bb}\\
R(t) &=& \int_0^t \left(-S^\prime(u)\right)F(t-u) du + R(0) + I(0)F(t), \label{eq:Cc}
\end{eqnarray}
with initial conditions $S(0)$, $I(0)$, $R(0)$.  
The parameter $\beta$ measures the transmission rate (also called infection rate \cite{Kirkeby2017,Weiss2013}) and the function $F(t)$ is the cumulative distribution of the time from infection to recovery.    
When $F(t)=1-e^{-\gamma t}$, the exponential distribution with mean $\gamma^{-1}$, our model reduces to the standard SIR model (see Murray \cite{Murray93} for example).  
For completeness, the proof of existence and uniqueness of the solution of System~(\ref{eq:Aa})-(\ref{eq:Cc}) is provided in the appendix. An alternative proof can be found in \cite{Hethcote80}.

The model parameters $\beta$ and $F(t)$ are epidemiologically relevant and provide insights into the outbreak. For example,  the basic reproductive number as defined by Lotka \cite{Heesterbeek2002,Lotka1911}:
\begin{equation} \label{eq:R0}
{\mathcal R}_0 = \int_0^\infty \frac{\beta}{N} S(t)(1-F(t)) dt 
\approx \frac{\beta}{\gamma},
\end{equation}
where $\gamma^{-1} = \int_0^\infty (1-F(t))dt$ is the average recovery time,  is  arguably the most widely used measure of the severity of an outbreak \cite{Heesterbeek96,Heffernan05}, at least in the absence of interventions to control it. It measures the expected number of secondary infections attributed to the index case in a na\"ive population. Other quantities of interest, such as the maximum number of infected individuals and the total number of infections, can be expressed in terms of the reproductive number $\mathcal{R}_0$, e.g. Weiss \cite{Weiss2013}.

For many diseases,  it is reasonable to assume that the disease progression from infection to recovery is known, either because the
disease is well characterized, or because the date of onset of symptoms, hospital admissions, and discharge data are available \cite{Who2014Ebola}.
Thus we will assume throughout this paper that we know the distribution of the  recovery period $F(t)$ and we will focus on estimating the transmission rate $\beta$.

\subsubsection*{Modeling the observed disease incidence} 

Let $\widetilde S(t)$, $\widetilde I(t)$ and $\widetilde R(t)$ denote the
observed number of susceptible, infected and recovered individuals 
as a function of time.  We make the following modeling assumptions:
\begin{enumerate} 
    \item[(A1)] The true underlying dynamics follows the SIR dynamics  described by Eqs~(\ref{eq:Aa})-(\ref{eq:Cc}) with known fixed population size
    $N$ and time-to-recovery distribution $F(t)$.
    \item[(A2)] A constant fraction $p$ of newly infected individuals is observed, that is $\widetilde S^\prime(t) = pS^\prime(t)$, with  $0<p<1$.  The same fraction $p$ of initial cases is
    observed, i.e., $\widetilde I(0) = pI(0)$, $\widetilde R(0)= p R(0)$, and $\widetilde S(0)= N- \widetilde I(0)- \widetilde R(0)$.
    \item[(A3)] The recovery distribution is the same for observed and unobserved infected individuals.
\end{enumerate}
Under these assumptions, the observed number of infected individuals at time $t$ is
\begin{equation}\label{eq:auxhI}
\widetilde I(t) = \int_0^t \left(-\widetilde S^\prime(u)\right)(1-F(t-u))du +\widetilde I(0)(1-F(t))=p I(t),
\end{equation}
and similarly, $\widetilde R(t) = p R(t)$.  The number of observed
susceptible individuals is 
\begin{equation}\label{eq:auxhS}
\widetilde S(t) = (1-p)N + pS(t).
\end{equation}
Eq~(\ref{eq:auxhS}) follows by solving the differential equation $\widetilde S'(t)=pS'(t)$ and using the
identity  $\widetilde S(0)=N(1-p)+pS(0)$, which results from (A2) and $N=S(0)+I(0)+R(0)$.

These equations 
capture the intuitive idea that under-reported incidence 
results in a larger number of observed susceptible and fewer infected and recovered individuals through the epidemic evolution. 
Consider the ratio, which yields from Assumption (A2) and Eqs~(\ref{eq:auxhI}) and (\ref{eq:auxhS}):
\begin{equation}\label{eq:ratio}
\frac{-\widetilde S^\prime(t)}{\widetilde S(t) \widetilde I(t)}= \frac{-pS^\prime(t)}{pI(t)\left[(1-p)N+pS(t)\right]}
= \frac{\beta}{N}  \frac{S(t)}{(1-p)N + pS(t)} 
=\frac{\beta}{N} v(t). 
\end{equation}
For a standard SIR model with $p=1$, the ratio $v(t)$ is unity.  However, for the observed process, the ratio $v(t)$ starts at one and then monotonically decreases over time.   It follows that  fitting an SIR model to observed incidence data,  neglecting the under-reporting, will produce a nearly unbiased, but possibly noisy estimate for $\beta$ early in the outbreak when $v(t) \approx 1$.  
As more data becomes available and $v(t)$ decreases, the estimated transmission
rate $\beta$  will under-estimate the true value.  As a consequence, one might at later times in an outbreak underestimate the severity of the outbreak and call the epidemic under control prematurely.

The following theorem describes the dynamics of the observed number of susceptible, infected and recovered individuals when only a fraction $p$  of the infected individuals are observed.

\begin{theorem} 
Under assumptions (A1), (A2), and (A3), the process of the observed individuals evolves according to the following set of integral-differential equations:
\begin{eqnarray}\label{theorem:A}
\widetilde S^{\prime}(t)&=&-\frac{\beta}{Np} \widetilde S(t) \widetilde I(t) + \frac{\beta(1-p)}{p} \widetilde I(t) \label{eq:pSIR.S}\\
\widetilde I(t)&=&\int_{0}^{t} \left(-\widetilde S'(u)\right)\left(1-F(t-u)\right)du+ \widetilde I(0)\left(1-F(t)\right)  \label{eq:pSIR.I}\\
\widetilde R(t)&=&\int_{0}^{t} \left(-\widetilde S'(u)\right)F(t-u) du + \widetilde R(0) + \widetilde I(0)F(t).
\label{eq:pSIR.R}
\end{eqnarray}
\end{theorem}

The conclusion of the theorem follows from algebraic manipulations of Eqs~(\ref{eq:Aa}) to (\ref{eq:auxhS}).
The addition of the positive term $((1-p)\beta/p) \widetilde I(t)$ to  $\widetilde S^\prime(t)$ implies a slower depletion rate of the observed susceptible population than would be expected under the standard SIR model.  Note that this positive term must be small enough such that $-\widetilde S^\prime(t)\geq0$, for all $t\geq0$  and all $p>0$, condition imposed from  Assumption (A2).
Assumption (A2) of observing the same fraction $p$ of initial infected and recovered individuals was established only for the technical mathematical proofs of Eq~(\ref{eq:auxhI}) and $\widetilde R(t) = p R(t)$.
This mathematical assumption will be relaxed in the following local dynamics definition.

\subsubsection*{A stochastic model for the observed incidence}

Observed incidences of disease are typically reported at regular time intervals.  
Precisely, let $0=t_0 < t_1 < t_2 < \dots < t_n$ denote the boundaries of the observation windows.  For simplicity,
we assume that $t_k = k \Delta$, and we denote by $Y_k$ the number of new cases of the disease observed in the interval $(t_{k-1}, t_k]$, $k=1,2,\dots, n$. We also assume that the new cases $Y_k$ depend on the actual
observed past history of incidences $\mathcal{H}_{k-1}=\{Y_1, Y_2,\ldots, Y_{k-1}\}$. As a result, our model takes into account the impact of  fluctuations in the reports. Indeed, imagine that 
the reported cases $Y_k$ are much larger than what is predicted by Model (\ref{eq:pSIR.S})-(\ref{eq:pSIR.R}).  That excess of cases will alter the observed dynamics of the outbreak, making it progress faster.  
Similarly, smaller numbers of incidences will slow down the outbreak.  
The following model takes into account past fluctuations in the incidence 
to model locally the dynamics of the process at each time interval given the 
past history.  

\begin{definition} \label{definition:exp}
Let $Y_1,Y_2,\ldots,Y_k$ be the sequence of observed incidences and
assume that the cumulative probability distribution $F$ for the
time to recovery is continuous.
We model the local dynamics of the observed number of susceptible $\widetilde S_k(t)$
and infected $\widetilde I_k(t)$ individuals at time $t$ in the interval $(t_{k-1},t_k]$ 
through the set of differential-integral 
equations
\begin{eqnarray}
\label{eq:pSIR.Sk}
\widetilde S_k^\prime(t) &=& -\frac{\beta}{Np} \widetilde S_k(t) \widetilde I_k(t) + \frac{\beta(1-p)}{p} \widetilde I_k(t) \\
\widetilde I_k(t) &=&\int_{t_{k-1}}^t (-\widetilde S'_k(u))\left(1-F(t-u)\right)du 
\label{eq:pSIR.Ik} \\
&& \qquad +\sum_{j=1}^{k-1} \frac{Y_j}{\Delta}\int_{t_{j-1}}^{t_j}(1-F(t-u))du+\widetilde I(0)\left(1-F(t)\right), \nonumber
\end{eqnarray}
with initial conditions $\widetilde S_k(t_{k-1}) = \widetilde S(0) - \sum_{j=1}^{k-1} Y_j$ and with the convention that $\sum_{j=1}^0 Y_j = 0$, where both $\widetilde S_k(t_{k-1})>0$ and $-\widetilde{S}'_k(t)\geq0$ for all $t\geq 0$ and $p>0$.
For this model, the conditional expectation of incidence given the past history
is
\begin{eqnarray}\label{eq:muk}
\mu_k={\mathbb E}[Y_k|Y_1,Y_2,\ldots,Y_{k-1}]=\int_{t_{k-1}}^{t_k} \frac{\beta}{Np} \widetilde S_k(u) \widetilde I_k(u) -
 \frac{\beta(1-p)}{p} \widetilde I_k(u) du, 
\end{eqnarray}
for all $k=1,2,\ldots,n$.
\end{definition}

\begin{remark}
Continuity of the cumulative distribution of the time to recovery
$F$ implies that $\widetilde I_k(t_k)$ is left continuous.
Furthermore, if $F$ has a probability density, then $\widetilde I_k(t)$ admits
a right-hand derivative at $t_{k-1}$.
\end{remark}

The local model described in Definition \ref{definition:exp} has the same infection dynamics, Eq~(\ref{eq:pSIR.Sk}), as the global model.  What differs is the evolution of the number of infected individuals, and how it relates
to the history of past incidences.  The following heuristic serves to  motivate Eq~(\ref{eq:pSIR.Ik}) in 
Definition \ref{definition:exp}.  Decompose the integral in 
Eq~(\ref{eq:pSIR.I})
for the number of infected individuals into a sum over each observed window
$(t_{j-1},t_j]$ to write
\begin{eqnarray*}
\widetilde I(t) &=&\int_{t_{k-1}}^t \left(-\widetilde S'(u)\right)\left(1-F(t-u)\right)du \\
&& \qquad + \sum_{j=1}^{k-1} \int_{t_{j-1}}^{t_j} \left(-\widetilde S'(u)\right)\left(1-F(t-u)\right)du  + \widetilde I(0)\left(1-F(t)\right).
\end{eqnarray*}
To get $\widetilde I_k(t)$, replace $\widetilde S^\prime(t)$ by its local instantiation 
$\widetilde S_k^\prime(t)$ on $(t_{k-1},t_k]$ and $\widetilde S^\prime(t)$ by 
$Y_j/\Delta$, the empirical rate of new infections, on the interval $(t_{j-1},t_j]$. 
This is interpreted as assuming that the $Y_j$ new infections in the interval $(t_{j-1},t_j]$
occur uniformly in that interval.   
This allows us to take into account the actual number of observed incidence in each time interval instead of using modeled derived quantities, which provides the needs flexibility for our local epidemic model to better 
track more complex epidemic 
dynamics than is possible using a global generalized SIR model.

\bigskip

We use the expression 
for the conditional expectation of incidences in the 
interval $(t_{k-1},t_k]$ given the time series of past observed incidences
in Definition \ref{definition:exp} to model the conditional distribution
of $Y_k$ given ${\mathcal H}_{k-1} = \{Y_1,Y_2,\ldots,Y_{k-1}\}$.  Specifically,
we assume that the conditional distribution of
$Y_k|{\mathcal H}_{k-1}$ is negative binomial
\begin{equation}\label{eq:NegBinom}
Y_k | Y_1,Y_2,\ldots,Y_{k-1} \sim \mbox{NegBinom}
\left ( \frac{\mu_k}{\mu_k+r} , r \right )
\end{equation}
with probability of success 
$\mu_k/(\mu_k+r)$ 
and shape parameter 
$r > 0$,
where
$\mu_k$ is the conditional expectation defined in Eq~(\ref{eq:muk}).
With this parametrization,
the conditional expectation and variance are
\begin{equation}
 {\mathbb E}[Y_k|{\mathcal H}_{k-1}]=\mu_k \qquad \text{and} \qquad {\mathbb V}[Y_k|{\mathcal H}_{k-1}] = {\mathbb E}[Y_k|{\mathcal H}_{k-1}] + \frac{\mu_k^2}{r},   
\end{equation}
respectively.  The shape parameter $r$
controls the amount of over dispersion when compared to a Poisson
distribution for which $\mathbb V[Y_k|{\mathcal H}_{k-1}]={\mathbb E}[Y_k|{\mathcal H}_{k-1}]$.  
In particular, as the shape parameter $r$ grows to infinity, the negative
binomial model converges to a Poisson distribution with rate $\mu_k$.
Thus, the negative binomial distribution allows us to account for the
extra-Poisson variability that arises in our model.  
Other distributions are possible, such as  beta negative binomial distribution \cite{Wang11} or the Conway-Maxwell-Poisson distribution \cite{Shmueli05}.  

\medskip

With repeated application of the chain rule, 
we combine the set of conditional distributions for $Y_k|Y_1,Y_2,\ldots,Y_{k-1}$
into a joint
likelihood for the model parameters
\begin{eqnarray} \label{eq:I}
L(\beta,p,r) &=& \prod_{k=2}^n {\mathbb P}[Y_k|Y_1,Y_2,\ldots,Y_{k-1}] \times {\mathbb P}[Y_1]\\
\label{eq:like}
&=&\prod_{k=2}^n \frac{\Gamma(y_k+r)}{\Gamma(r)\Gamma(y_k)}\times \left(\frac{\mu_k}{\mu_k+r}\right)^{y_k}\left(\frac{r}{\mu_k+r}\right)^{r} \times {\mathbb P}[Y_1]
\end{eqnarray}
where $\Gamma$ denotes the gamma function and $\mu_k$ depends only on $\beta$ and $p$. 
Since, in the model formulation, the distribution of $Y_1$ does not contain any information about the transmission rate and the fraction of observed cases, the term $\mathbb{P}[Y_1]$ is dropped from the likelihood. 

\subsubsection*{Approximation of the conditional expectation}

To reduce the computational burden required to numerically solve 
the set of differential-integral equations (\ref{eq:pSIR.Sk})-(\ref{eq:pSIR.Ik}),
and the ensuing integration in Eq~(\ref{eq:muk}) to evaluate the 
conditional expectation,  we propose to approximate the conditional 
expectation $\mu_k$ by  linearizing both $\widetilde S_k(u)$
and $\widetilde I_k(u)$ around $t_{k-1}$ in Eq~(\ref{eq:muk}), and 
integrate the result explicitly. The following lemma
encapsulates the resulting approximation.

\begin{lemma}\label{lemma:UniformApprox} 
Assume that the cumulative probability distribution $F$ for the time to recovery has a probability density $f$. 
The conditional expectation $\mu_k$ can be approximated by
\begin{equation}\label{eq:mukApprox}
    \mu_k=\max \left ( -\Delta \widetilde S_{k-1}^\prime\left[1+\frac{\Delta}{2}\left(\frac{ \widetilde I_{k-1}^\prime}{\widetilde I_{k-1}}-\frac{\beta}{p}\frac{\widetilde I_{k-1}}{N}\right)-\frac{\beta\Delta^2}{3p}\frac{ \widetilde I_{k-1}^\prime}{N}\right],
    0 \right ),
\end{equation}
when $\widetilde I_{k-1}\neq 0$ and ${\widetilde S_{k-1}}/{N}\geq(1-p)$, and  $\mu_k=0$ otherwise. Here,
\begin{eqnarray}
\label{eq:pSIR.sk}
\widetilde S_{k-1}=\widetilde S_k(t_{k-1})&=&\widetilde S(0) - \sum_{j=1}^{k-1} Y_j
\\
\label{eq:pikU}
\widetilde I_{k-1}=\widetilde I_k(t_{k-1})&=&\sum_{j=1}^{k-1} \frac{Y_j}{\Delta}\int^{t_{j}}_{t_{j-1}}\left(1-F(t_{k-1}-u)\right)du+\widetilde I(0)\left(1-F(t_{k-1})\right)
\\
\label{eq:pdskU}
\widetilde
 S_{k-1}^\prime=\widetilde S^\prime_k(t_{k-1})&=&-\frac{\beta}{p}\left(\frac{\widetilde S_{k-1}}{N}-(1-p)\right)\widetilde I_{k-1}
\\
\label{eq:pdikU}
\widetilde I'_{k-1} = \widetilde I^\prime_k(t_{k-1})&=&-\widetilde S'_{k-1}-\sum_{j=1}^{k-1}\frac{Y_j}{\Delta}\left( F(t_{k-j})-F(t_{k-j-1})\right)
-\widetilde I(0)f(t_{k-1}),
\end{eqnarray}
for all $k=1,2,\ldots,n$.
\end{lemma}

\subsubsection*{Proof of Lemma \ref{lemma:UniformApprox}}

Eqs~(\ref{eq:pSIR.sk})-(\ref{eq:pdikU}) follow directly from the definition of $\widetilde S(t_{k-1})$ and the evaluation of Eqs~(\ref{eq:pSIR.Sk})-(\ref{eq:pSIR.Ik}) at $t_{k-1}$. To prove Eq~(\ref{eq:pdikU}), we first take the derivative of $\widetilde I_k(t)$, Eq~(\ref{eq:pSIR.Ik}), with respect to $t$ and simplify it as follows:
\begin{eqnarray*}
\widetilde I'_{k}(t)=-\widetilde S'_k(t)-\sum_{j=1}^{k-1}\frac{Y_j}{\Delta}\int_{t_{j-1}}^{t_j}f(t-u)du+\int_{t_{k-1}}^t\widetilde S_k'(u)f(t-u)du-\widetilde I(0)f(t)\hspace{.6cm}\\
=-\widetilde S'_k(t)-\sum_{j=1}^{k-1}\frac{Y_j}{\Delta}(F(t-t_{j-1})-F(t-t_j))+\int_{t_{k-1}}^t\widetilde S'_k(u)f(t-u)du-\widetilde I(0)f(t).
\end{eqnarray*}
Then, we evaluate at $t_{k-1}$ and simplify the resulting equation, using the definition of each $t_k=\Delta k$:
\begin{eqnarray*}
\widetilde I'_{k}(t_{k-1})&=&-\widetilde S'_k(t_{k-1})-\sum_{j=1}^{k-1}\frac{Y_j}{\Delta}(F(t_{k-1}-t_{j-1})-F(t_{k-1}-t_j))-\widetilde I(0)f(t_{k-1})\\
&=&-\widetilde S'_k(t_{k-1})-\sum_{j=1}^{k-1}\frac{Y_j}{\Delta}(F(t_{k-j})-F(t_{k-j-1}))-\widetilde I(0)f(t_{k-1}).
\end{eqnarray*}
From the definition, both $\widetilde S_{k-1}$ and $\widetilde I_{k-1}$ are non-negative quantities, and the hypothesis ${\widetilde S_{k-1}}/{N}\geq(1-p)$ implies that $-\widetilde S'_{k-1}\geq 0$, 
for all $p>0$. Therefore, all these equations are well defined.
In the proof of Eq~(\ref{eq:mukApprox}), the
linear approximation of both $\widetilde S_k(u)$
and $\widetilde I_k(u)$ around $t_{k-1}$ are:
\begin{eqnarray}
\widetilde S_k(u) &\approx& \widetilde S_k(t_{k-1}) + (u-t_{k-1}) \widetilde S_k^\prime(t_{k-1}) \\
\widetilde I_k(u) &\approx& \widetilde I_k(t_{k-1}) + (u-t_{k-1}) \widetilde I_k^\prime(t_{k-1}).
\end{eqnarray}
Substituting these equations in the integrand of Eq~(\ref{eq:muk}) and solving it yields:
\begin{eqnarray*}
\mu_k&\approx&\frac{\beta}{Np}\left(\Delta \widetilde S_{k-1}\widetilde I_{k-1}+\frac{\Delta^2}{2}(\widetilde S'_{k-1}\widetilde I_{k-1}+\widetilde I'_{k-1} \widetilde S_{k-1})+\frac{\Delta^3}{3}\widetilde I'_{k-1} \widetilde S'_{k-1}\right)\\
&&-\frac{\beta (1-p)}{p}\left(\Delta\widetilde I_{k-1}+\frac{\Delta^2}{2}\widetilde I'_{k-1}\right)\\
&=&\Delta\left(\frac{\beta}{Np}\widetilde S_{k-1}\widetilde I_{k-1}-\frac{\beta (1-p)}{p}\widetilde I_{k-1} \right)+\frac{\beta\Delta^3}{3Np} \widetilde S'_{k-1}\widetilde I'_{k-1}\\
&&+ \frac{\Delta^2}{2}\left(\widetilde I'_{k-1}\left(\frac{\beta}{Np}\widetilde S_{k-1}-\frac{\beta (1-p)}{p}  \right)+\frac{\beta}{Np}\widetilde S'_{k-1}\widetilde I_{k-1}\right).
\end{eqnarray*}
When $\widetilde I_{k-1}=0$, from Eqs~(\ref{eq:pikU})  and (\ref{eq:pdikU}),
$\widetilde S'_{k-1}=0$ and $\widetilde I'_{k-1} =0$. Then  $\mu_k\approx0$.  When $\widetilde I_{k-1}\neq 0$,
using the definition of $\widetilde S'_{k-1}$ in the previous equation and simplifying it yields:
\begin{eqnarray*}
\mu_k&\approx&-\Delta \widetilde S'_{k-1}+\frac{\beta\Delta^3}{3Np} \widetilde S'_{k-1}\widetilde I'_{k-1} - \frac{\Delta^2}{2}\left(\frac{\widetilde I'_{k-1}}{\widetilde I_{k-1}} \widetilde S'_{k-1} -\frac{\beta}{Np}\widetilde S'_{k-1}\widetilde I_{k-1}\right),
\end{eqnarray*}
where the conclusion of Eq~(\ref{eq:mukApprox}) follows.

\begin{remark}
Better approximations for $\mu_k$ can be obtained using
higher order Taylor expansions for $\widetilde S_k(u)$
and $\widetilde I_k(u)$.  This requires the distribution $F$ of 
time to recovery to have higher order derivatives.
\end{remark}

\subsubsection*{Identifiability}

It is known that the measured growth rates in early SIR outbreaks are insensitive to
under-reporting.  Indeed, in early outbreaks, $S(t) \approx N$ and hence
$I^\prime(t) \approx (\beta-\gamma) I(t)$.  Under Assumption (A2), we have that
$\widetilde I(t) = pI(t)$ and $\widetilde I^\prime(t) = p I^\prime(t)$,  which 
imply that
\[
\frac{d}{dt} \log I(t) = \frac{d}{dt} \log \widetilde I(t) = \beta-\gamma.
\]
It follows that the disease incidence grows exponentially with rate 
$\beta-\gamma$, irrespective on the fraction $p$ of observed incidence. Hence the transmission rate $\beta$ can be estimated if the
recovery rate $\gamma$ is known, but the  fraction $p$
cannot be estimated at that early stage of the outbreak.

As the outbreak matures and moves away from its early exponential growth 
phase, it becomes possible to estimate both the transmission rate $\beta$ 
and the fraction $p$ of observed cases.
The following theorem provides verifiable
conditions for both these parameters to be identifiable.

\begin{theorem} \label{theorem:C}
Set 
\begin{eqnarray}
U_k &=& \int_{t_{k-1}}^{t_k} \frac{\widetilde  S_k(u)}{N} \widetilde I_k(u) du\\
 V_k &=& \int_{t_{k-1}}^{t_k} \widetilde I_k(u) du.
\end{eqnarray}
If the vector $(U_1,U_2,\ldots,U_m)$ and $(V_1,V_2,\ldots,V_m)$
are linearly independent, then $\beta$ and $p$ are identifiable.
\end{theorem}
The proof of Theorem \ref{theorem:C} is found in the appendix.

\begin{remark}
As we note earlier, $\beta$ and and $p$ are not identifiable 
in the early stages of an outbreak.  This is also evident in
Theorem \ref{theorem:C}:  In the early stages, 
we have that $\widetilde S_k(u) \approx N$, so that the vectors
$(U_1,\ldots,U_k)$ and $(V_1,\ldots,V_k)$ are essentially co-linear.
Later in the outbreak, as $S_k(u)$ is no longer close to $N$, both parameters
become identifiable.
\end{remark}

\subsection*{Bayesian parameter estimation}

We use the Metropolis-Hastings algorithm to draw Monte-Carlo Markov chain (MCMC) \cite{Makowski02} samples from the posterior distribution of the model parameters given the epidemic outbreak data.  Our implementation transforms the original parameters $\Theta=(\beta,p,r)$ into $\widetilde \Theta= (\xi,\eta,\rho) \in {\mathbb R}^3$, where $\xi=\log(\beta)$,  $\eta=\log(p/(1-p))$ and $\rho=\log(r)$, and selects proposals from a multivariate Gaussian distribution ${\mathcal N}(\cdot|\widetilde\Theta_{m})=\mathrm{N}(\widetilde\Theta_{m},\Sigma),$ with mean $\widetilde\Theta_{m}$ and diagonal covariance matrix $\Sigma$  with entries $0.001$, $0.01$, and $0.01$.  The results presented in the next section are from 40,000 MCMC samples gathered after 40,000 burn-in iterations when starting from $\Theta_0=(0.5,0.5,25)$.  Our implementation used the approximation for $\mu_k$ presented in Lemma \ref{lemma:UniformApprox}.  

\begin{table}[!ht]
\begin{center}
\small{\caption{
{\bf Metadata of analyzed data sets.}}
\begin{tabular}{|l|l|l|l|l|l|}
\hline
 Country  & Initial reports  & Intervention & Population&$n$& $I(0)$\\
\hline
USA       &January 20 &March 22& 331,002,651 & 485 &5 \\
Brazil    &February 26 &March 24& 212,559,417& 448& 5\\
Mexico    &February 28 &March 23& 128,932,753& 446& 6\\ 
Argentina & March 03&March 19& 45,195,774    & 442& 5\\
Chile     &March 03&March 24 & 19,116,201    & 442& 5\\
Colombia  &March 06&March 25 & 50,882,891    & 439&5\\ 
Peru      &March 07&March 16& 32,971,854     & 438&9\\
Panama    &March 10&March 24 & 4, 314,767    & 435& 5\\
\hline
\end{tabular}
\label{Table:Data}}
\end{center}
\end{table}

Following \cite{Ferretti20,LiGuan20}, we model the distribution of time to recovery from COVID-19 as the convolution of a lognormal distribution (with mean=5.2 and  sdlog=0.662) with a Weibull distribution (with mean=5 and  sd=1.9). The mean and standard error of the resulting recovery time distribution are 10.27 and 4.32, respectively.   We refer the interested reader to \cite{Bar20} for a detailed description of additional disease progression parameters of  SARS-CoV-2 infection.

Separate chains were run for the time series of incidence data from each country, using all the data from the date of the first confirmed COVID-19 cases to May 18th, 2021 (see Table~\ref{Table:Data}). The assumption of a constant transmission rate does not hold, as each country implemented various mitigation and control strategies, from national lockdown orders to closing of public meeting places (see Table~\ref{Table:Data} which shows the date on first implementation of mitigation as reported in \cite{Wikipedia}).  To avoid having to model the change in the transmission rate resulting from the implementation of mitigations, our parameter estimation starts on the first day of intervention as reported in Table \ref{Table:Data}. We still use the whole time series from the time of first confirmed incidence to estimate the number of infected individuals as defined by Eq~(\ref{eq:pSIR.Ik}). 

To reduce the impact of weekly reporting patterns (e.g. fewer cases are reported over the weekend) we apply a moving average of seven days to the raw incidence counts before executing the  MCMC algorithm.  Finally, the initial conditions $\widetilde{S}_0=N-pI(0)-pR(0)$, $R(0)=0$,  $\widetilde{I}_0=pI(0)$ are set using the reported national population counts and number of initial cases as reported in Table~\ref{Table:Data}.

\section*{Results and discussion}

\subsection*{Analysis of COVID-19 incidence data}

\begin{figure}[!ht]
\begin{center}
\includegraphics[width=1\textwidth]{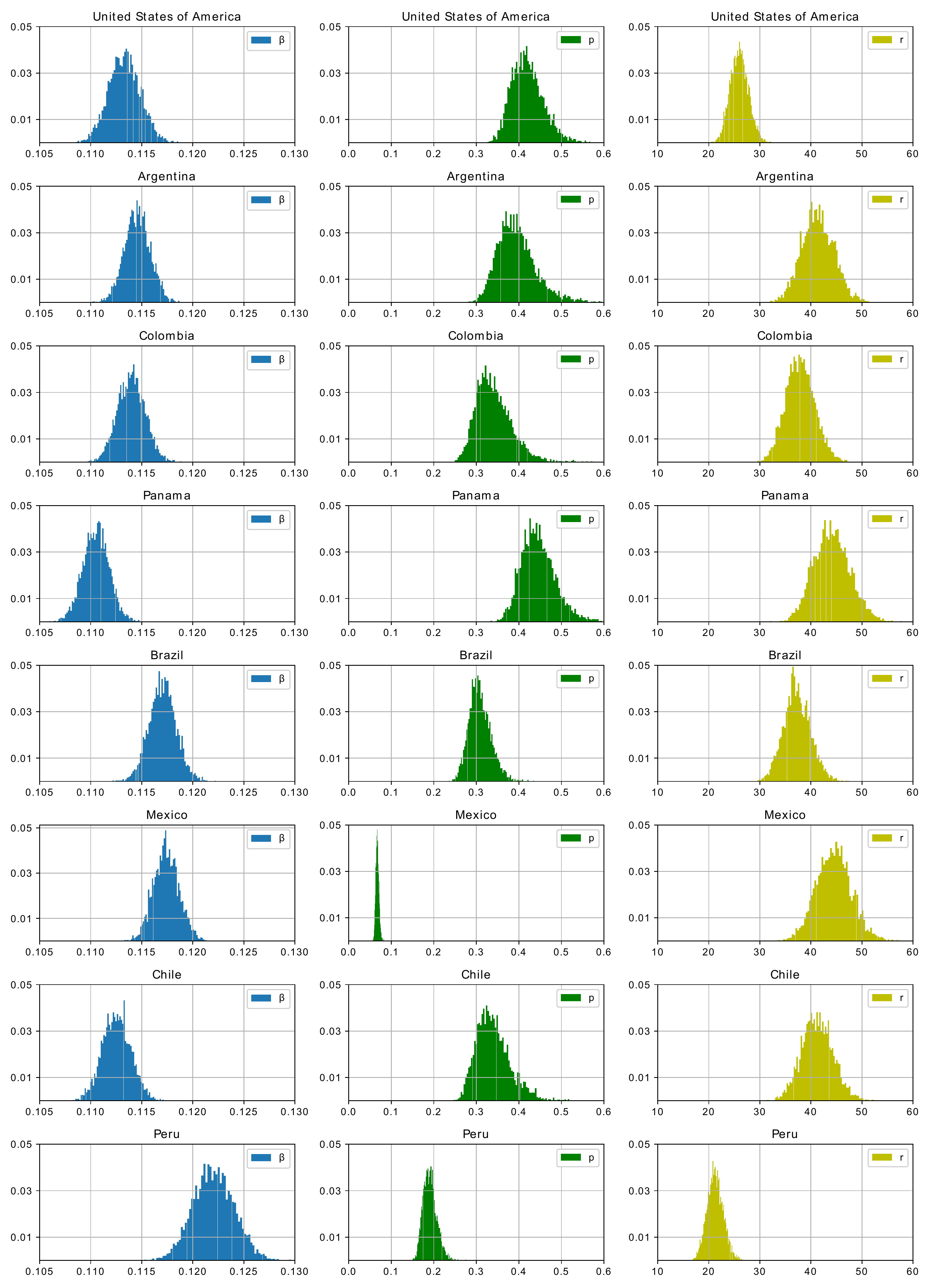}
\end{center}
\caption{{\bf Histograms of the marginal posterior distribution of the transmission rate $\beta$ (left), fraction of reported cases $p$  (middle) and the negative binomial shape parameter $r$ (right) for each country.} The $x$-axis corresponds to the estimated values, and the
$y$-axis is the bin’s relative frequency.
\label{fig:Histo}}
\end{figure}

We performed separate Bayesian inferences for eight  American Countries: the United States of America (USA), Brazil, Mexico, Argentina, Chile, Colombia, Peru, and Panama.  Fig~\ref{fig:Histo}  shows histograms of the marginal posterior distribution of the transmission rate $\beta$ after the start of mitigation, the fraction observed $p$, and the negative binomial shape parameter $r$  for each country.  The median and 95\% confidence intervals of these posterior distributions are presented in Table \ref{Table:Summary}.

\begin{table}[!ht]
\centering
\small{\caption{
{\bf Parameter Median Values and Confidence Interval Estimations.
}}
\begin{tabular}{|l|l|l|l|l|l|l|l|l|}
\hline
 Country  & $\beta$ & $95\%$ CI  ($\beta$) & $p$& $95\%$ CI ($p$) &$r$ &$95\%$ CI ($r$) \\
\hline
USA      & $0.113$            &$[0.110, 0.116]$ & $0.418$         & $[0.359, 0.502]$&$25.990$ &$[22.728, 29.629]$   \\
Brazil    &$0.117$             &$[ 0.114, 0.120]$ & $0.307$         & $[0.264, 0.365]$&$37.133$ & $[32.372, 42.646]$ \\
Mexico      &$0.117$             &$[0.115, 0.120]$ & $0.067$         & $[0.061, 0.076]$&$44.260$ &$[37.896, 51.527]$  \\ 
Argentina        &$0.115$             &$[0.112, 0.117]$ & $0.390$         & $[ 0.324, 0.509]$& $41.259$ & $[35.451, 47.574]$  \\
Chile      &$0.112$             &$[0.110, 0.115]$ & $0.335$         & $[0.278, 0.431]$& $41.222$ & $[35.459, 47.478]$  \\
Colombia       &$0.114$             &$[0.111, 0.117]$ &  $0.336$         & $[0.277, 0.440]$&$ 37.920$ & $[32.705, 43.725]$ \\ 
Peru      &$0.122$             &$[0.118, 0.126]$ & $ 0.190$         & $[0.164, 0.226]$&$21.229$ & $[18.304, 24.525]$ \\
Panama      &$0.110$             &$[0.108, 0.113]$ & $0.443$         &$[0.379, 0.537]$& $44.003$ & $[37.691, 51.221]$ \\
\hline
\end{tabular}
\label{Table:Summary}}
\end{table}
Even though each country used different mitigation strategies, with various level of enforcement, the confidence intervals for  the transmission parameter of each of the eight countries overlap, with the
exception of Peru.  There are several hypothesis for why this
 may be the case:  the effectiveness of the various mitigation strategies is
 compromised by having a small fraction of non-compliant individuals, or
 most of the benefits of the mitigation strategy are achieved by wearing face
 masks and moderate social distancing.  A third hypothesis 
 is that the estimated transmission rate in 
 our model is a time average of the instantaneous transmission rates, and  
 that averaging lessens the differences in transmission rates.

Similarly, the posterior distributions for the fraction of observed incidence  are similar across most of the analyzed countries.  The two exceptions are Peru and Mexico, with the under-reporting in Mexico being particularly acute.  This is consistent with the observation that Mexico has one of lowest numbers of tests performed per reported case \cite{DailyTest}. While an under-reporting factor of about 15 is very large,  we believe this effect is real because of how well the model fits the data (see the appendix) and narrowness of the posterior distribution. 

Related analyses of COVID-19 data in Mexico have  used values for the fraction of reported cases of  $p=0.2$ or  $p=0.4$ to analyze and forecast the evolution of the COVID-19 pandemic and hospital demands \cite{Capistran21,Saldana20}. These values are closer to the values that we found for the other Latin American countries. However, these values were not derived from the data. It would be interesting to use our model to investigate the under-reporting in Mexico at a county level to see how the results would differ from local to national levels. 

Excess deaths \cite{cuellar2021} provide an alternative measure of the true impact of COVID-19.  Using that measure, \cite{chowell2021b} reports that COVID-19 deaths in Mexico are under-reported by a factor of 3, whereas we show a factor of 15 for under-reported incidence.  This difference may be due differential testing rates of deceased and infected individuals which may arise from the standard of care of severely ill patients admitted to intensive care units that requires COVID-19 testing \cite{Murthy_2020,Capistran21}.

Our analysis flags Peru as being different from the other countries
both in term of having a higher transmission rate, and a lower reported
fraction.  Our analysis does not reveal why this is the case, and further 
analysis incorporating country level explanatory variables to predict
transmission rates and under-reporting is needed to uncover the reasons why Peru 
is different from the other countries in America we studied.

Finally, the estimate of the shape parameter $r$ of the negative binomial distribution shows that the relative inflation of the  Poisson variance ranges from 2\%-5\%.  That effect is statistically significant. Again, the distributions across the eight countries are commensurate, with the United States and Peru exhibiting more extra Poisson variability than the other countries.

\subsection*{Under-estimation of the transmission rate}
In light of Eq~(\ref{eq:ratio}), we suggested in the introduction
that failing to account for under-reporting leads 
to underestimating the transmission rate $\beta$.  Here, we numerically
demonstrate this effect by
fitting an SIR-type model directly to raw incidence data, deliberately neglecting  
to model under-reporting.  
The median and 95\% confidence intervals of 
the posterior distribution for the transmission rate when modeling 
under-reporting, and when not are displayed in   Table~\ref{Table:SummaryUnderEstimation}.

\begin{table}[!ht]
\small{\caption{ {\bf Parameter Median Values and Confidence Interval Estimations for the observed $\beta_p$ and true underlying $\beta_1$ rates.}}\label{Table:SummaryUnderEstimation}
	\begin{center}
		\begin{tabular}{|l|l|l|l|l|}
			\hline
			Country  & $\beta_p$&  $95\%$ CI  ($\beta_p$) &$\beta_1$ & $95\%$ CI  ($\beta_1$) \\
			\hline
			USA      & $0.113$        &$[0.110, 0.116]$           & $0.107$            &$[0.105, 0.109]$   \\
			Brazil    &$0.117$        &$[ 0.114, 0.120]$         &$0.110$             &$[0.108, 0.111]$   \\
			Mexico      &$0.117$      &$[0.115, 0.120]$          &$0.105$             &$[0.103, 0.106]$ \\ 
			Argentina        &$0.115$ &$[0.112, 0.117]$      &$0.110$             &$[0.108, 0.112]$  \\
			Chile      &$0.112$   &$[ 0.110, 0.115]$         &$0.107$ & $[0.105, 0.108]$ \\
			Colombia       &$0.114$      &$[0.111, 0.117]$        &   $0.109$            & $[0.107, 0.110]$ \\
			Peru      &$0.122$     &$[0.118, 0.126]$          &$0.109$             &$[0.107, 0.112]$ \\
			Panama      &$0.110$       &$[0.108, 0.113]$     &$0.105$             &$[0.103, 0.106]$  \\
			\hline
		\end{tabular}     
	\end{center}}
\end{table}

The parameter $\beta_p$ coincides with the values from Table~\ref{Table:Summary}, while $\beta_1$ refers to estimates when the fraction
observed is $p=1$.  The posterior distribution for shape parameter $r$ were similar for $p$ unknown and $p$ fixed.

Observe that in all cases, the 95\% confidence intervals for the transmission rate do not 
overlap.  This shows that knowledge of the fraction of reported incidence is statistically
important.

\subsection*{Variation on the fraction of reported cases}

In this section, we consider modeling and estimating a time dependent fraction $p(t)$ of reported incidence, which can arise from uneven availability of COVID-19 tests \cite{Capistran21,Rothe20,Wu2020}. To this end, we model the reported fraction $p(t)$ with a piece-wise constant function:
\begin{equation}\label{eq:p(t)}
    p(t)= \sum_{k=1}^{M}p_{k}\mathbb{I}_{[\xi_{k-1},\xi_{k})}(t),
\end{equation}
for all $0<t\leq t_n < \xi_M$, where  $\mathbb{I}_{[\xi_{k-1},\xi_{k})}$ denotes
the indicator function for each interval $[\xi_{k-1},\xi_k)$.
We regularize the sequence of reported fractions $p_1, p_2,\ldots,p_M$ by  adding the the penalty 
\begin{equation}\label{eq:penalty}
\frac{M-1}{2}\ln(\lambda)-\frac{\lambda}{2} \, \sum_{k=2}^M (p_k-p_{k-1})^2 -\frac{M-1}{2}\ln(2\pi)
\end{equation}
to the loglikelihood.  We assume that the variation between reported fraction, $p_k-p_{k-1}$, are 
identically and independently normally distributed with mean zero and variance $1/\lambda$.

\begin{figure}[ht!]
\centering 
\includegraphics[width=.87\textwidth]{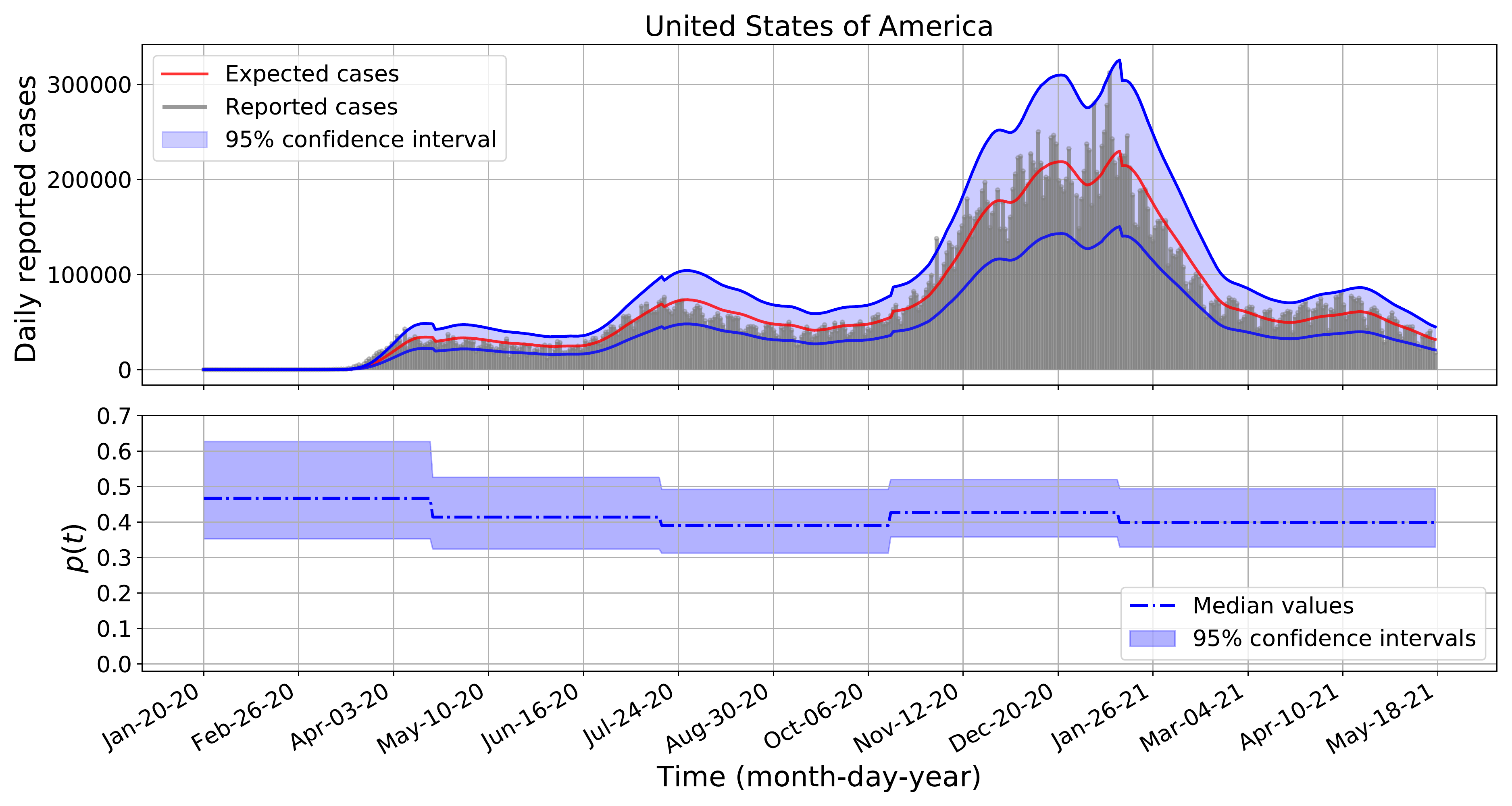}
\caption{{\bf Daily COVID-19  incidence and fraction of reported cases for the United States of America from January 20, 2020  to May 18, 2021.}}
\label{fig:USA_p_varying}
\end{figure}

\begin{figure}[ht!]
\centering 
\includegraphics[width=.87\textwidth]{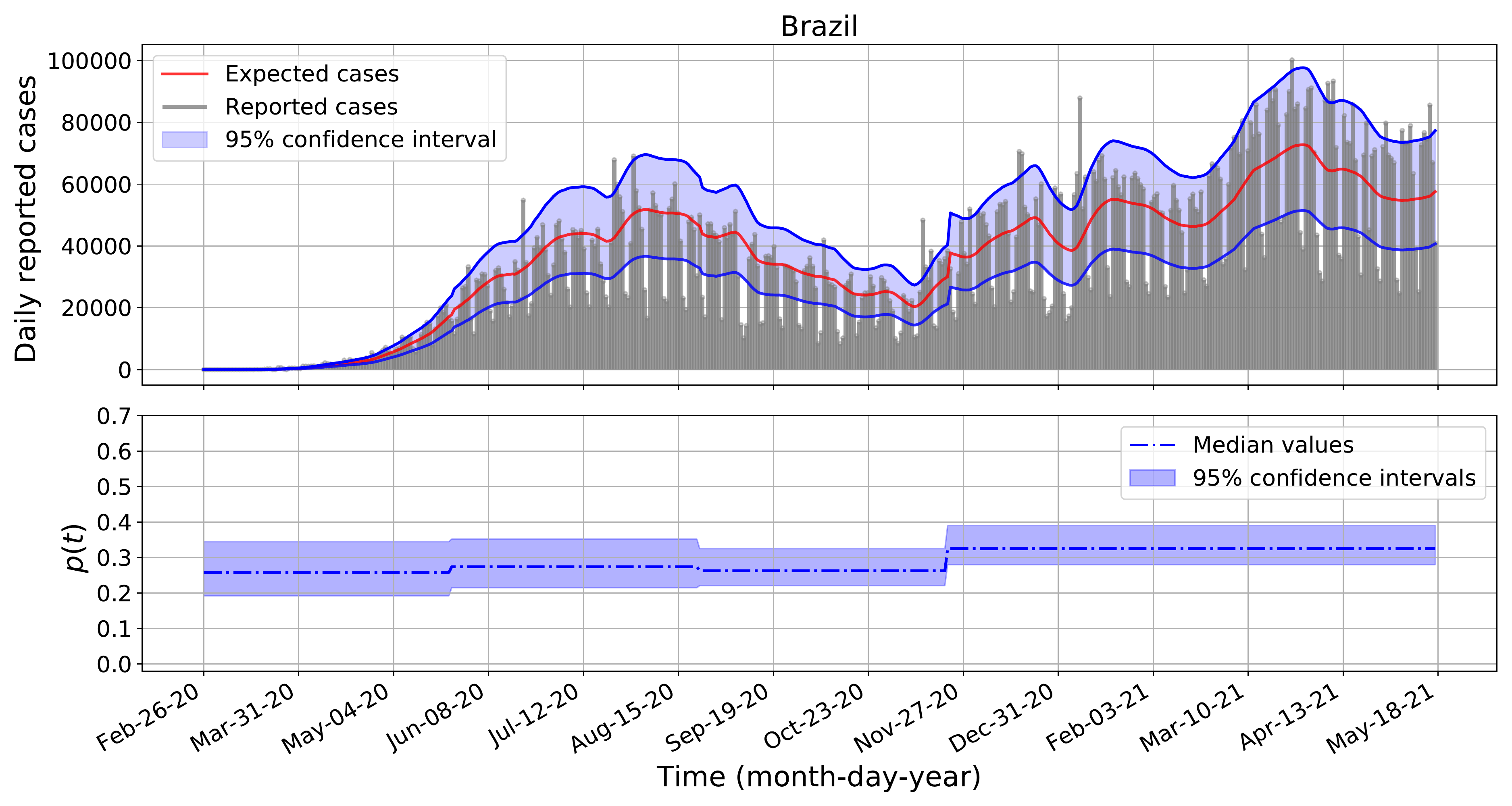}
\caption{{\bf Daily COVID-19 incidence and fraction of reported cases for Brazil from  February 26, 2020 to May 18, 2021.}}
\label{fig:BR_p_varying}
\end{figure}

\begin{figure}[ht!]
\centering 
\includegraphics[width=.87\textwidth]{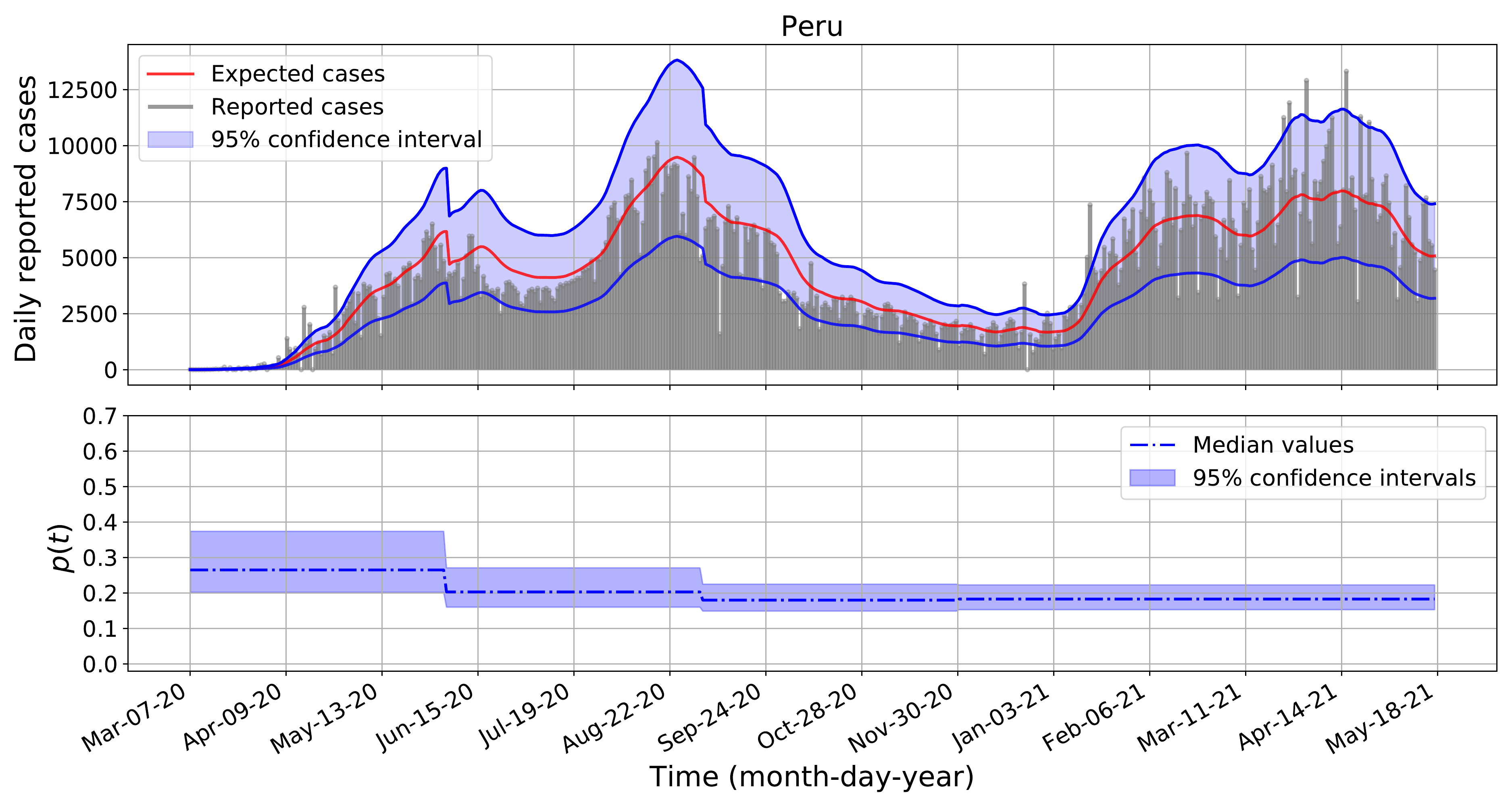}
\caption{{\bf Daily  COVID-19  incidence and fraction of reported cases for Peru from March 07, 2020  to May 18, 2021.}}
\label{fig:PR_p_varying}
\end{figure}

Similarly as in the previous section,
we performed separate Bayesian inferences 
to estimate the posterior distributions of $\beta$, $p_1, p_2, \dots, p_{M}$, $r$ and $\lambda$ for each analyzed country: the United States of America, Brazil, Mexico, Argentina, Chile, Colombia, Peru, and Panama. 
We defined $p(t)$ with constant pieces  of length modulo 90 days and we use the equations from Lemma~\ref{lemma:UniformApprox_pk} to compute the expected incidence $\mu_k$.
These equations generalize the equations
from Lemma~\ref{lemma:UniformApprox} when $p=p_k$ for all  $k=1,2,\ldots,M$, 
see the appendix for further details.
The median and 95\% confidence intervals of the posterior distributions of $\beta$, $r$, and $\lambda$
are presented in Table \ref{Table:p_varying}. 
For clarity in the presented results for $\lambda$ values, we decided to round them to the nearest integer values.  The
analogous results for the posterior distributions
for each reported fraction, $p_1, p_2, p_3, p_4$ and $p_5$, are plotted in the second panel of
Figs~\ref{fig:USA_p_varying}-\ref{fig:PR_p_varying} for the United States of America, Brazil, and Peru. The corresponding results
for Mexico, Argentina, Chile, Colombia, and Panama are shown in  the second panel of Figs~\ref{fig:MX_p_varying}-\ref{fig:PN_p_varying} in the appendix.
In all cases, the $95\%$ confidence intervals for each $p_k$ values are displayed in the blue-shadow areas, while their median values are  plotted in blue-dashed-dotted lines.
\begin{table}[!ht]
\centering
\small{\caption{{\bf Parameter Median Values and Confidence Interval Estimations varying the $p$ values.}}
\small{\begin{tabular}{|l|l|l|l|l|l|l|l|l|l|l|}
\hline
 Country  & $\beta$ & $95\%$ CI  ($\beta$)  &$r$ &$95\%$ CI ($r$)&$\lambda$&$95\%$ CI ($\lambda$)\\
\hline
USA      & $0.114$            &$[0.111, 0.117]$      &$26.259$ &$[22.833, 29.993]$ &$231$&$[25, 1843]$\\
Brazil    &$0.118$             &$[0.115, 0.120]$       &$37.876$ & $[32.864, 43.406]$ 
&$230$&$[19, 1905]$\\
Mexico      &$0.119$             &$[0.116, 0.121]$       &$51.400$ &$[44.576, 59.145]$ 
&$722$&$[51, 2798]$\\ 
Argentina        &$0.114$       &$[0.111, 0.116]$      & $41.818$ & $[35.933, 48.632]$ &$642$& $[79, 2056]$ \\
Chile      &$0.113$             &$[0.110, 0.116]$    &   $41.232$ & $[35.137, 48.253]$
&$786$ &$[86, 2870]$\\
Colombia       &$0.114$      &$[0.112, 0.117]$      & $38.553$ & $[33.022, 44.653]$ &$364$&$[18, 2561]$\\ 
Peru      &$0.122$             &$[0.118, 0.125]$       &$22.189$ & $[19.138, 25.786]$ &$416$ &$[32, 1262]$ \\
Panama      &$0.110$             &$[0.108, 0.113]$      & $46.731$ & $[39.783, 54.376]$ &$152$ &$[30, 812]$ \\
\hline
\end{tabular}}
\label{Table:p_varying}}
\end{table}	

Additionally, in the first panel of Figs~\ref{fig:USA_p_varying}-\ref{fig:PN_p_varying} we show the confidence interval of the model estimates for the daily COVID-19 incidence for each country, where
the expected median of reported cases, $\mu_k$, are plotted in red lines, the upper and lower predicted bounds are plotted in blue lines, while the expected incidences lie in the blue-shadow area with  probability of $95\%$. The negative binomial distribution function, Eq~(\ref{eq:NegBinom}), was used to build the confidence bounds.
To estimate the expected cases, the parameter values for $\beta$ and $r$  were set equal to the values provided in Table~\ref{Table:p_varying} and the $p_k$ values were set to the estimated median values of $p(t)$ as shown in the second panel of each figure and for each country, respectively.

From Table~\ref{Table:p_varying}, the marginal posterior distributions for the parameter $\lambda$ overlap for all analyzed country. All these marginal posterior 
distributions skewed to the right with large values. For most countries, the confidence bands for $p(t)$ include a constant function. That is, statistically, we do not have enough evidence to reject the hypothesis that the reported fraction $p(t)$ for each country is not a constant function during the entire analyzed data set.  And for countries that have a small variance $\lambda^{-1}$ for the increment $p_k-p_{k-1}$, we have further evidence that $p(t)$ is nearly constant. The one country for which a constant $p(t)$ is not retained is Mexico (see Fig~\ref{fig:MX_p_varying}).

The second panels of  Figs~\ref{fig:USA_p_varying}-\ref{fig:PR_p_varying} shows that there are some variations across all $p_k$ confidence intervals for the United States of America, Brazil, and Panama. Interestingly, the confidence intervals of $p_k$ for each country are all contained in a wider confidence band than those obtained when assuming a constant fraction $p$
of observed cases as reported in Table~\ref{Table:Summary}.
Comparing Table~\ref{Table:Summary} and Table~\ref{Table:p_varying}, we  see that there are not significant changes on the posterior distributions for $\beta$ and $r$ when we assume $p$ constant and $p$ variable.  In general, we observe more variation for the observed proportion $p$ across the countries than within a country.  The latter result is not surprising, as countries implemented different testing policies which may affect the way the incidence data were reported \cite{Wu2020,Capistran21,DailyTest}.

\subsection*{Strengths and weaknesses of the proposed local SIR model}

Our model locally exploits the SIR dynamics, using past observations to 
set the initial conditions.  This results in a flexible model that can
fit complex patterns, such as multiple waves that typically require
a time varying transmission rate, with a single parameter.  This flexibility
comes at a cost:  our single estimated transmission rate is a time average of the 
true time varying one.  And while we show that our model empirically 
fits the data well within the confidence bounds, 
we  over-estimate the expected incidence in the valleys and under-estimate near the peaks.
It follows that the derived estimates for the reproductive number near a
local bottom of an outbreak will have a positive bias, leading to a 
more conservative view of the effect of mitigation.

Our formulation can be generalized to build epidemic models having non-parametric  transmission
rates.  Such models will alleviate the weakness discussed above, and can be used to identify model-based uncertainties in models.  These extensions will be presented in a forthcoming paper. We are also planning to extend the model by incorporating the exposed class, which will provide a more realistic model to study COVID-19 pandemic. As  COVID-19 disease progression depends on both the length of time an individual remains in the exposed and infectious classes \cite{Bar20}. This model extension would help us to analyze the effect of different infectious period distributions  that could change at the early outbreak due to interventions such as testing, isolation or contact tracing.

Finally, our model has a limited ability to estimate time-varying under-count  fractions.  Numerical experiments have shown that adding more 
flexibility to how the latter varies over time degrades our ability to estimate the transmission rate.

\section*{Conclusion}

We present a new extension of the standard SIR epidemiological models to study the under-reported incidence of infectious diseases. The new model reveals that fitting a SIR model type directly to raw incidence data will under-estimate the true infectious rate when neglecting  under-reported  cases. Using the epidemic model we also present a Bayesian methodology to estimate the transmission rate and  fraction of under-reported incidence with confidence bounds that result directly from incidence data. We also argue that our statistical model can properly track and estimate complex incidence reports, where the resulted estimates update as more data are incorporated.

Using our methodology on the COVID-19 example, we found that the confidence intervals for the transmission rates overlap across the eighth analyzed American countries: the United States of America, Brazil, Argentina, Chile, Colombia, Peru, and Panama.  In all the cases, the median transmission rates are above 0.105 and below 0.122 (see Tables~\ref{Table:Summary}, \ref{Table:SummaryUnderEstimation}, and \ref{Table:p_varying}).  And, for most countries, the confidence bands for  the time dependent fraction of reported cases $p(t)$ include a constant function, and they also provide a range values for the fraction of reported cases per each country. In average, from January 03, 2020 to May 18, 2021: the reported incidence fraction for the United States of America and Panama varies from 0.3 to 0.6; the reported incidence fraction for Brazil, Chile, Colombia, and Argentina varies from 0.2 to 0.5; the reported incidence fraction for Peru varies from 0.15 to 0.35 while for Mexico varies from 0.05 to 0.1 
(see Figs \ref{fig:USA_p_varying}-\ref{fig:PN_p_varying}).

\appendix
\section*{Appendix}

\subsection*{Proof of existence and uniqueness of solutions of the generalized SIR model}
To prove existence and uniqueness of solutions of System (\ref{eq:Aa})-(\ref{eq:Cc}), it is further assumed that the fraction of recovered individuals is defined through a probability distribution function, $F:[0,\infty)\rightarrow [0,1]$, with the following properties.  
\begin{property}\label{P1} 
    There exists an integrable function
    $f:[0,\infty)\rightarrow [0,\infty)$ such that
    $$F(t)= \int_0^t f(u) du \quad \text{and}\quad 
    \int_0^\infty f(u) du=1,$$
    for all  $t\in [0,\infty)$.
\end{property}
\begin{property} \label{P2}
The average recovery time  is finite, i.e.,  $$\frac{1}{\gamma}=\int_0^\infty (1-F(t))dt<\infty.$$
\end{property}
\begin{theorem}\label{Thm:ExistanceAndUniq} Let $U$ be an open set of $[0,N] \times [0,N] \times [0,N]$ $\times [0,\infty)$ and $K$ a compact subset of \, $U$ containing $(S(0), I(0), R(0),$ $t_0)$, the initial condition of System~(\ref{eq:Aa})-(\ref{eq:Cc}), with $f(t)$ continuously differentiable with respect to $t$, $t\geq0$ in $U$. Then there exists a unique solution of System~(\ref{eq:Aa})-(\ref{eq:Cc}) through the point $(S(0),I(0),R(0))$ at $t=0$, denoted $X(S(t),I(t),R(t),t)$, with 
$X(S(0),I(0),R(0),(0)) = (S(0),I(0),R(0))$, for all $t$ such that  $X(S(t),I(t),R(t),t)\in K$.
\end{theorem}

\subsubsection*{Proof of Theorem \ref{Thm:ExistanceAndUniq}}
From Property~\ref{P2}, 
System (\ref{eq:Aa})-(\ref{eq:Cc}) is well defined and it is equivalent to
\begin{eqnarray}
S^\prime(t) &=& -\frac{\beta}{N} S(t)I(t)\hspace{4.5cm} \label{eq:dA}\\
I^\prime(t) &=& \int_0^t f(t-u)S^\prime(u)du -S^\prime(t)-I(0) f(t) \label{eq:dB}\\
R^\prime(t) &=& -\int_0^t f(t-u)S^\prime(u)du  + I(0)f(t), \label{eq:dC}
\end{eqnarray}
which is obtained by taking the derivative with respect to $t$ of Eqs~(\ref{eq:Bb}) and (\ref{eq:Cc}) and using Property~\ref{P1}. 
Therefore, it is enough to prove existence and uniqueness of solutions of 
System~(\ref{eq:dA})-(\ref{eq:dC}).
It follows that
the function $G:U\rightarrow \mathbf{R}^3$ defined by 
\begin{equation}\label{eq:ODEs}
G(S,I,R,t)=(S^\prime(t),I^\prime(t),R^\prime(t))    
\end{equation}
is continuously differentiable in $U$, see for example \cite[pp. 32]{Wiggins03}. Since $\frac{\partial G}{\partial S}$,  $\frac{\partial G}{\partial I}$, $\frac{\partial G}{\partial R}$, and $\frac{\partial G}{\partial t}$ exist and are continuous in $U$, then $G$ is continuously differentiable in $U$. Therefore, the solution of System (\ref{eq:dA})-(\ref{eq:dC}) exists for the initial condition $S(0)$, $I(0)$, $R(0)$ and  is unique in $K$.

\subsection*{Proof of Theorem \ref{theorem:C}}
Set $\alpha_1 = \beta/p$ and $\alpha_2 = \beta(1-p)/p$.  Since  
\[  
p = 1 - \frac{\alpha_2}{\alpha_1} \quad \mbox{and} \quad \beta = \alpha_1 - \alpha_2,
\]
identifiability of $\alpha_1$ and $\alpha_2$ implies identifiability of $\beta$ and $p$.
We can estimate $\alpha_1$ and
$\alpha_2$ by minimizing the sum of squares
\begin{equation}
\sum_{k=1}^m \left ( Y_k - \alpha_1 U_k - \alpha_2 V_k \right )^2.
\end{equation}
The two parameters are identifiable if and only if the vectors $(U_1,\ldots,U_m)$
and $(V_1,\ldots,V_m)$ are not co-linear.

\subsection*{Modeling the time dependence fraction of reported incidence}

The following definition describes the dynamics of the observed susceptible and infected individuals when  
constant fractions $p_k$ of infected individuals are observed at each interval $(t_{k-1},t_{k}]$,  i.e., 
$\widetilde{S}_k'(t)=p_k{S}'(t)$ for all $t$ in that interval of time. 
This hypothesis allows us to study the case when the parameter $p$ is a piece-wise time dependent function, as it is defined in Eq~(\ref{eq:p(t)}).

\begin{definition}\label{Def:Local_p_varying}
Let $Y_1,Y_2,\ldots,Y_k$ be the sequence of observed incidences and 
assume that the cumulative probability distribution $F$ for the time to recovery is continuous. We model the local dynamics of the observed number of susceptible $\widetilde S_k(t)$ and infected $\widetilde I_k(t)$ individuals at time $t$ in the interval $(t_{k-1},t_k]$  through the set of differential-integral  equations:
\begin{eqnarray}\label{eq:AppendixBa}
\widetilde S_k^\prime(t) &=& -\frac{\beta}{Np_k} \widetilde S_k(t) \widetilde I_k(t) +  \beta\widetilde I_k(t)\left(\frac{1-p_k}{p_k}+\sum_{j=1}^{k-1}\frac{Y_j(p_k-p_j)}{Np_k\,p_j}+\frac{I(0)(p_k-p_1)}{Np_k}\right)\hspace{.5cm} \\\nonumber
\widetilde I_k(t) &=&\int_{t_{k-1}}^t \left(-\widetilde S'_k(u)\right)\left(1-F(t-u)\right)du
 +p_k \sum_{j=1}^{k-1}\frac{Y_j}{\Delta p_j} \int_{t_{j-1}}^{t_j} \left(1-F(t-u)\right)du\\\label{eq:AppendixBb}
 &+&p_k I(0)\left(1-F(t)\right),
\end{eqnarray}
with initial conditions for the observed  susceptible individuals
\begin{equation}\label{eq:AppendixBc}
    \widetilde S_{k}(t_{k-1})=N-p_1 I(0) - \sum_{j=1}^{k-1} Y_j
\end{equation}
and under the hypothesis $1\geq p_k>0$,
$\widetilde S_k(t_{k-1})>0$, and  $-\widetilde{S}'_k(t)\geq0$  for all $t$ and   $k=1,2,\ldots,n$.
For this model, the conditional expectation of incidence given the past history
is
\begin{eqnarray}\label{eq:AppendixBd}
\mu_k={\mathbb E}[Y_k|Y_1,Y_2,\ldots,Y_{k-1}]=\int_{t_{k-1}}^{t_k} (-\widetilde S'_k(u))du,
\end{eqnarray}
for all $k=1,2,\ldots,n$.
\end{definition}

Note that Definition~\ref{definition:exp} and  Definition~\ref{Def:Local_p_varying} 
are the same when $p=p_k$ for all $k=1,\ldots,n$.
In the following, we provide the mathematical motivation  of Definition~\ref{Def:Local_p_varying}, using similar ideas as from the derivation of Definition~\ref{definition:exp}.

First, from the definition of $I(t)$, we re-write Eq~(\ref{eq:Bb})
as follows:
\begin{eqnarray*}
I(t)&=&I(0)\left(1-F(t)\right)
 + \sum_{j=1}^{k-1}\frac{1}{p_j} \int_{t_{j-1}}^{t_j}(-p_j S'(u)) \left(1-F(t-u)\right)du \\
 &+&\frac{1}{p_k}\int_{t_{k-1}}^t \left(- p_k S'(u)\right)\left(1-F(t-u)\right)du.
\end{eqnarray*} 
Similarly for $S(t)$,
\begin{eqnarray*}
S(t)  &=&S(0)-\int_{0}^{t}(-S'(u))du\\
 &=&N-I(0)- \sum_{j=1}^{k-1}\frac{1}{p_j} \int_{t_{j-1}}^{t_j}(-p_jS'(u))du
 -\frac{1}{p_k}\int_{t_{k-1}}^t \left(- p_k S'(u)\right)du,
\end{eqnarray*} 
where in the second equation we used the hypothesis $N=S(0)+I(0)$.
Then, from the above two equations, we estimate  $I(t)$ and $S(t)$  with the  equations:
\begin{eqnarray}\label{Def:Ihat}
\widehat I(t)  &=&I(0)\left(1-F(t)\right) + \sum_{j=1}^{k-1}\frac{Y_j}{ \Delta p_j} \int_{t_{j-1}}^{t_j} \left(1-F(t-u)\right)du\\ \nonumber
&+&\frac{1}{p_k}\int_{t_{k-1}}^t \left(-\widetilde S_k'(u)\right)\left(1-F(t-u)\right)du\\\label{Def:Shat}
\widehat S(t)  &=&N-I(0)-  \sum_{j=1}^{k-1}\frac{Y_j}{p_j}-\frac{1}{p_k}\int_{t_{k-1}}^t \left(-\widetilde S_k'(u)\right)du, 
\end{eqnarray} 
which follow by estimating $p_jS'(u)$ with $\widetilde S_j(u)$ and then setting 
$\widetilde S_j(u)=Y_j/\Delta$ for all $u\in(t_{j-1},t_{j}]$ and all $j=1,2,\ldots,k-1$.
The last equality follows by assuming that the total cases $Y_j$ occur uniformly in the observed interval. 
Now, solving the integral of Eq~(\ref{Def:Shat}), with
the initial conditions $\widetilde{S}_k(t_{k-1})$ defined by Eq~(\ref{eq:AppendixBc}) and simplifying it, yields: 
\begin{eqnarray*}
\widehat S(t)&=&N-I(0)-  \sum_{j=1}^{k-1}\frac{Y_j}{p_j}-\frac{1}{p_k}\left(\widetilde S_k(t_{k-1})-\widetilde S_k(t) \right)\\
&=&\frac{1}{p_k}\widetilde S_k(t)+ N\left(1-\frac{1}{p_k}\right)-\sum_{j=1}^{k-1} Y_j\left(\frac{1}{p_j}-\frac{1}{p_k}\right)-I(0)\left(1-\frac{p_1}{p_k}\right).
\end{eqnarray*} 
The above equation implies that $\widetilde{S}'_{k}(t)=p_k\widehat{S}'(t)$. Therefore, from the estimates $\widehat I(t)$ and $\widehat S(t)$, Eqs~(\ref{Def:Ihat}) and (\ref{Def:Shat}), and the true transmission dynamics process, Eq~(\ref{eq:Aa}), we have:
\begin{eqnarray*}
&&\hspace{-0.9cm}\widetilde{S}'_k(t)=p_k\widehat{S}'(t)
\approx -p_k \frac{\beta}{N}\widehat{S}(t)\widehat I(t)
\\
&&=- \frac{\beta}{N}\left(\frac{1}{p_k}\widetilde S_k(t)+ N\left(1-\frac{1}{p_k}\right)-\sum_{j=1}^{k-1} Y_j\left(\frac{1}{p_j}-\frac{1}{p_k}\right)-I(0)\left(1-\frac{p_1}{p_k}\right)\right)\widetilde I_k(t)
\end{eqnarray*} 
where $\widetilde I_{k}(t)=p_k\widehat{I}(t)$. Therefore,  $\widetilde S_{k}(t)$ and $\widetilde I_{k}(t)$ satisfy Definition~\ref{Def:Local_p_varying}. 

\bigskip

The next lemma provides a recursive formula to approximate the conditional 
expectation $\mu_k$ defined by Eq~(\ref{eq:AppendixBd}). The equation results directly from solving the 
integral of Eq~(\ref{eq:AppendixBd}) with the linear approximation of both $\widetilde S_k(u)$
and $\widetilde I_k(u)$ around $t_{k-1}$. 
Its proof  is similar to the proof of Lemma~\ref{lemma:UniformApprox}.
\begin{lemma}\label{lemma:UniformApprox_pk} 
Assume that the cumulative probability distribution $F$ for the
time to recovery has a probability density $f$. 
The conditional expectation $\mu_k$ can be approximated by
\begin{equation}\label{eq:mukApprox2}
    \mu_k=\max \left ( -\Delta \widetilde S_{k-1}^\prime\left[1+\frac{\Delta}{2}\left(\frac{ \widetilde I_{k-1}^\prime}{\widetilde I_{k-1}}-\frac{\beta}{p_k}\frac{\widetilde I_{k-1}}{N}\right)-\frac{\beta\Delta^2}{3p_k}\frac{ \widetilde I_{k-1}^\prime}{N}\right],0 \right ),
    \end{equation}
when $\widetilde I_{k-1}\neq 0$ and $-\widetilde S'_{k-1}>0$, and  $\mu_k=0$ otherwise. Here,
\begin{eqnarray}
\label{eq:pkSIR.s}
\widetilde S_{k-1}&=&N-p_1 I(0) - \sum_{j=1}^{k-1} Y_j
\\
\label{eq:pkSIR.i}
\widetilde I_{k-1}&=&\frac{p_k}{\Delta}\sum_{j=1}^{k-1} \frac{Y_j}{p_j}\int^{t_{j}}_{t_{j-1}}\left(1-F(t_{k-1}-u)\right)du+ p_kI(0)\left(1-F(t_{k-1})\right)\hspace{0.5cm}
\\
\label{eq:pkSIR.ds}
\widetilde
 S_{k-1}^\prime&=&-\frac{\beta}{p_k}\left(\frac{\widetilde S_{k-1}}{N}-(1-p_k)-\sum_{j=1}^{k-1}\frac{Y_j(p_k-p_j)}{N p_j}-\frac{I(0)(p_k-p_1)}{N}\right)\widetilde I_{k-1}
\\
\label{eq:pkSIR.di}
\widetilde I'_{k-1}&=&-\widetilde S'_{k-1}-\frac{p_k}{\Delta}\sum_{j=1}^{k-1}\frac{Y_j}{p_j}\left( F(t_{k-j})-F(t_{k-j-1})\right)-p_k I(0)f(t_{k-1}),
\end{eqnarray}
for all $k=1,2,\ldots,n$.
\end{lemma}

\subsection*{Complementary numerical simulations}
In this section, we present a set of numerical simulations to complement the Results and discussion section. 

\begin{figure}[ht!]
\centering 
\includegraphics[width=.87\textwidth]{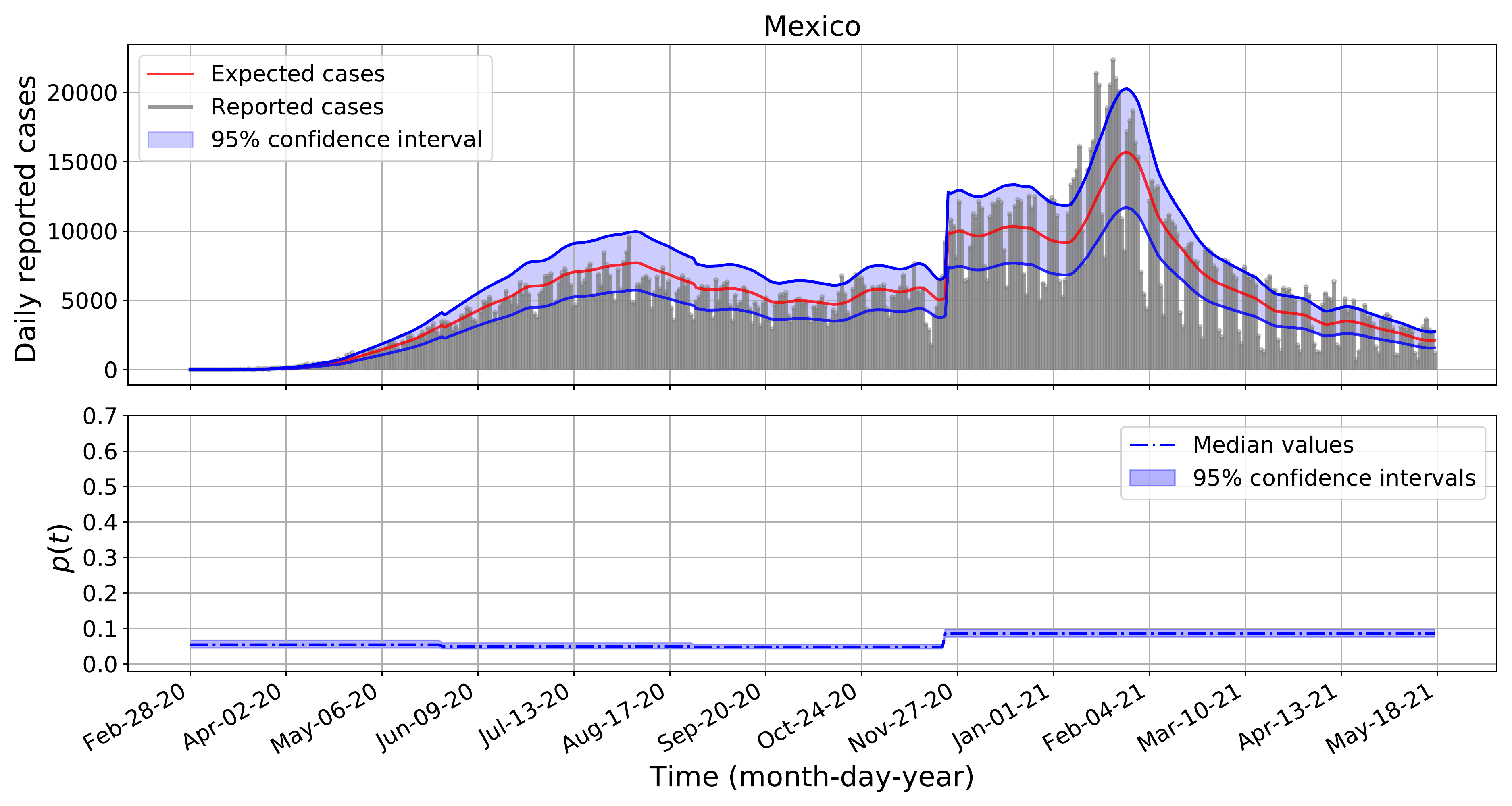}
\caption{{\bf  Daily  COVID-19 incidence and  fraction of reported cases for Mexico from February 28, 2020  to May 18, 2021.}}
\label{fig:MX_p_varying}
\end{figure}

\begin{figure}[ht!]
\centering 
\includegraphics[width=.87\textwidth]{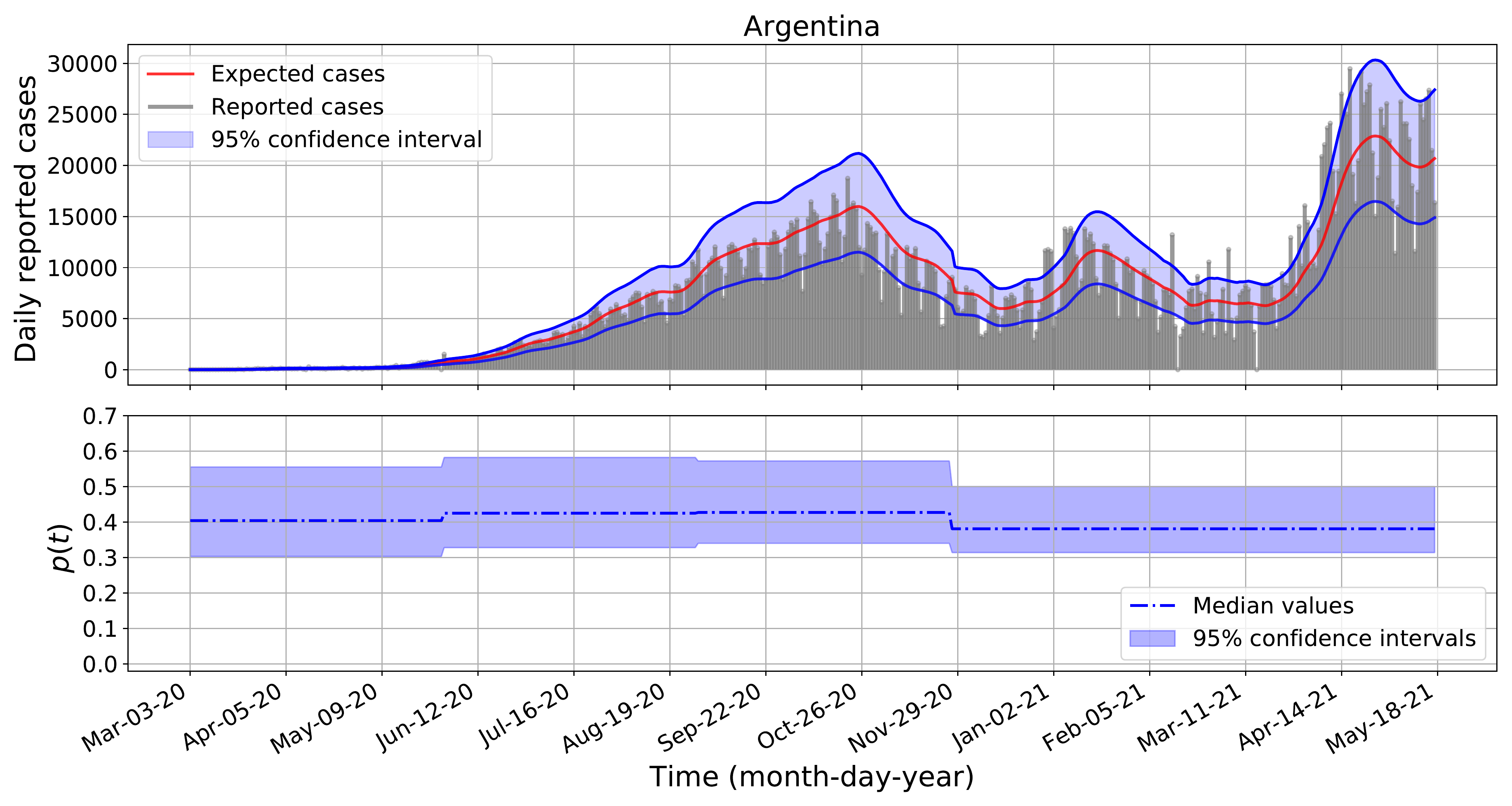}
\caption{{\bf Daily  COVID-19 incidence and  fraction of reported cases for Argentina from  March 03, 2020  to May 18, 2021.}}
\label{fig:AR_p_varying}
\end{figure}

\begin{figure}[ht!]
\centering 
\includegraphics[width=.87\textwidth]{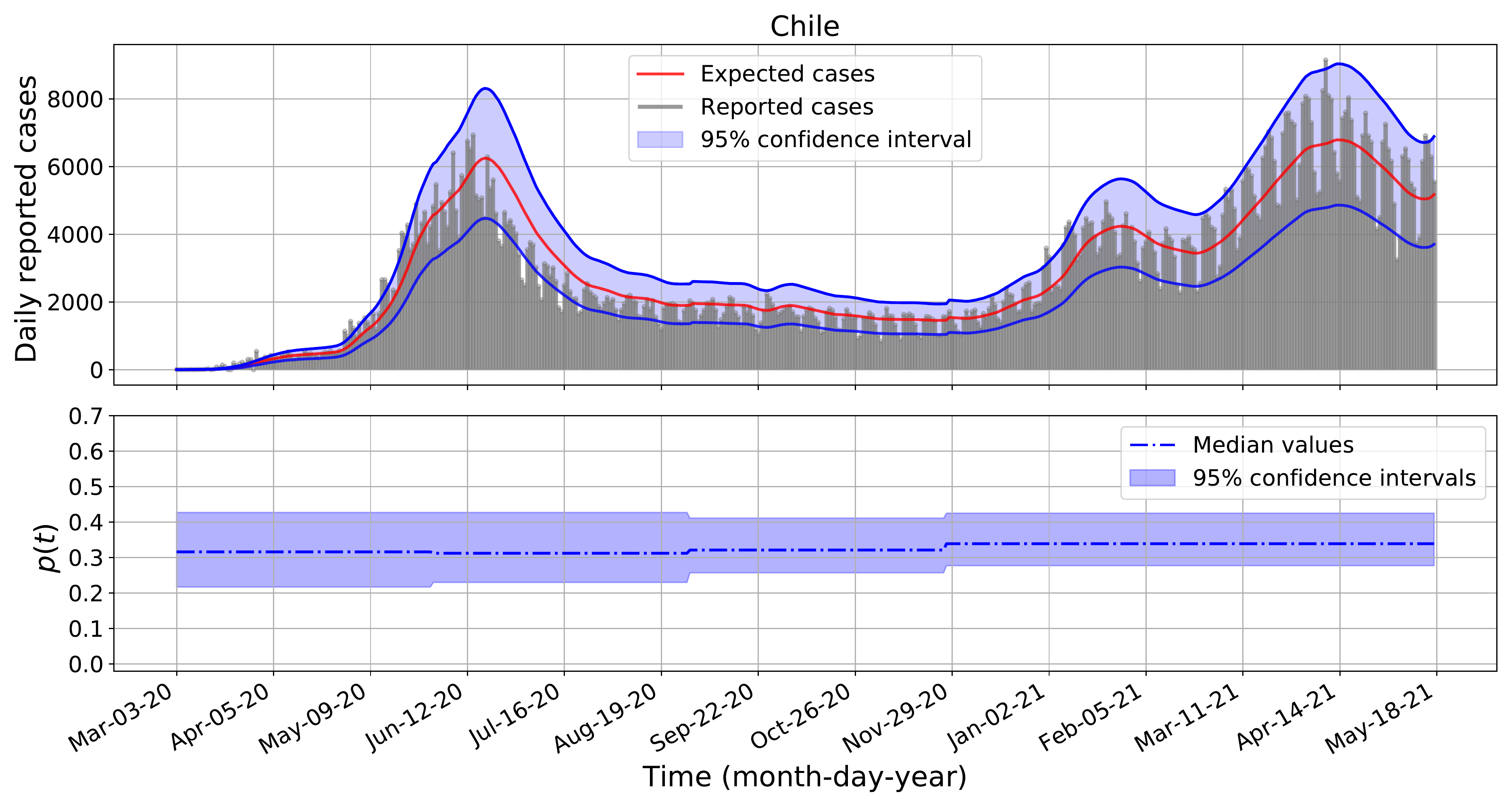}
\caption{{\bf Daily  COVID-19 incidence and  fraction of reported cases  for Chile from March 03, 2020  to May 18, 2021.}}
\label{fig:CH_p_varying}
\end{figure}

\begin{figure}[ht!]
\centering 
\includegraphics[width=.87\textwidth]{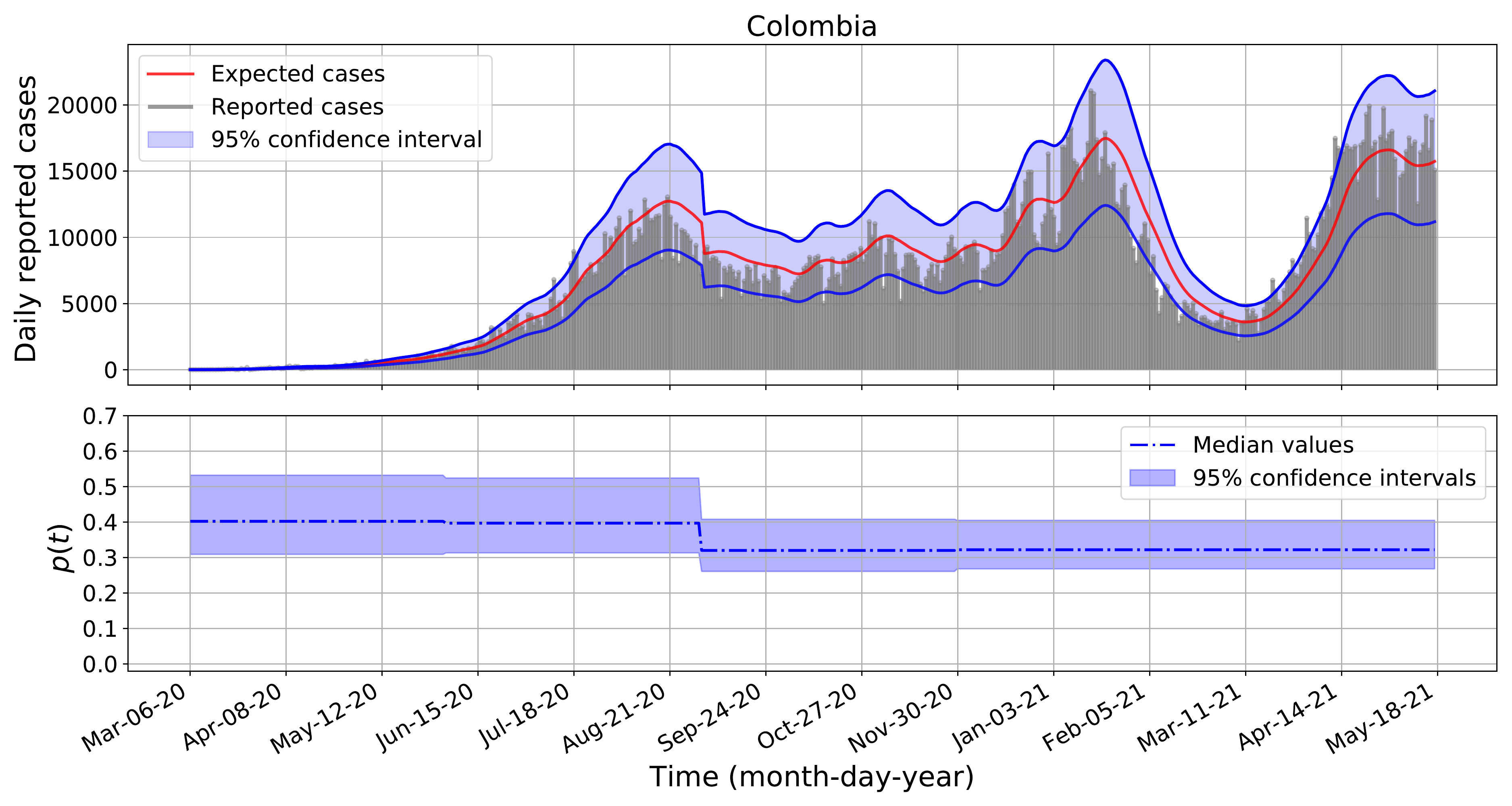}
\caption{{\bf Daily  COVID-19 incidence and  fraction of reported cases for Colombia from March 06, 2020  to May 18, 2021.}}\label{fig:CL_p_varying}
\end{figure}

\begin{figure}[ht!]
\centering 
\includegraphics[width=.87\textwidth]{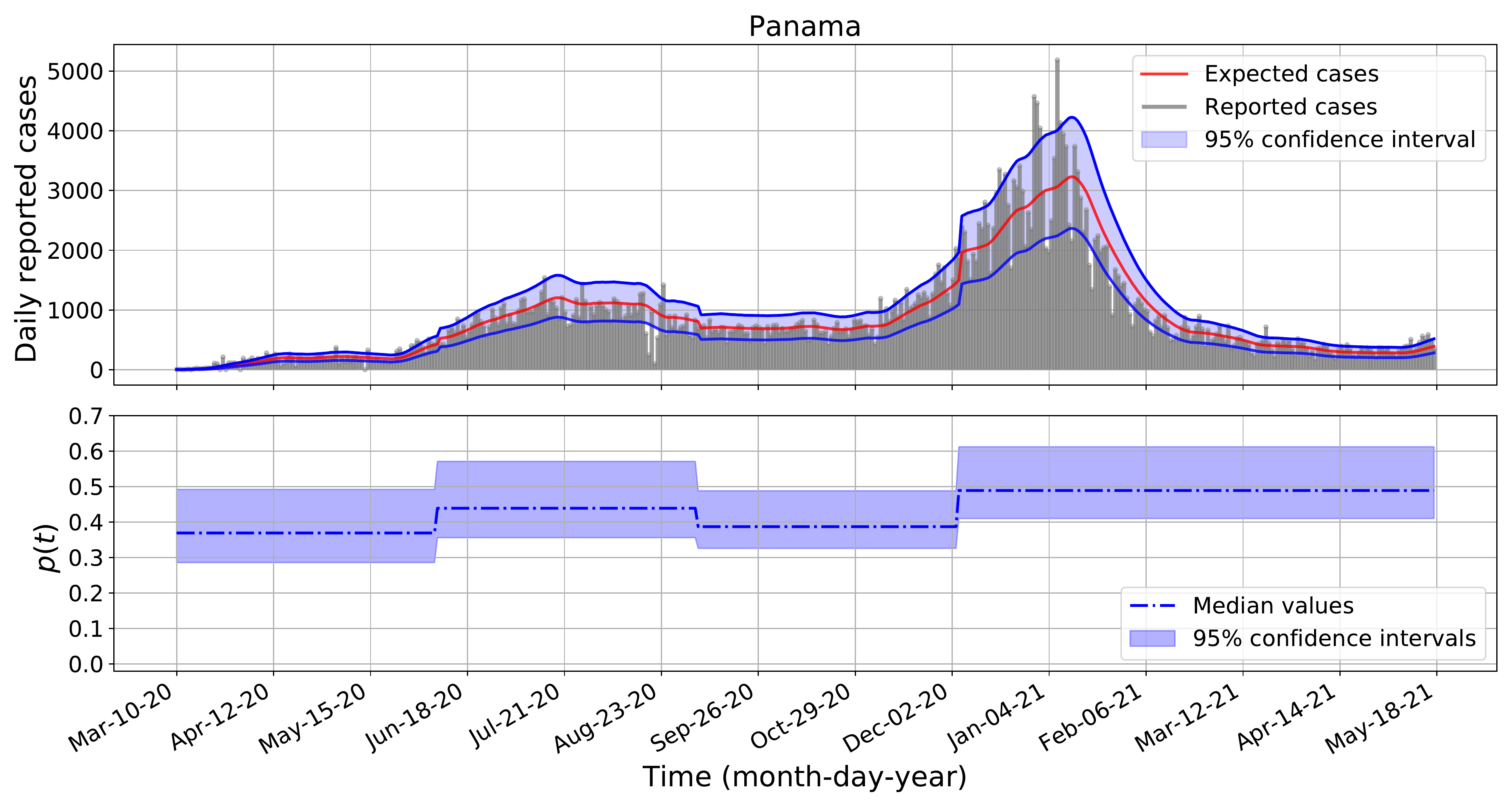}
\caption{{\bf Daily  COVID-19 incidence and fraction of reported cases for Panama from March 10, 2020  to May 18, 2021.}}
\label{fig:PN_p_varying}
\end{figure}

\end{document}